# Role of water in physics of blood and cerebrospinal fluid

Alexander Kholmanskiy

Science Center "Bemcom," Shenkursky Proezd, 11, 127340 Moscow, Russia

allexhol@ya.ru, http://orcid.org/0000-0001-8738-0189

**Abstract**

Known physical mechanisms of temperature dependence anomalies of water properties were used to explain the regularities in temperature dependence (TDs) of dynamic, electrical and optical characteristics of biological systems. The dynamics of hydrogen bonds in bulk and hydrated water affected the activation energies TDs of ion currents of voltage-dependent channels that regulate signaling and trophic bonds in the neuropil of the cortical parenchyma. The physics of minimizing the TD of the isobaric heat capacity of water made it possible to explain the stabilization and functional optimization of the thermodynamics of eyeball fluids at 34.5 °C and the human brain during sleep at 36.5 °C. At these temperatures, the thermoreceptors of the cornea and the cells of the ganglionic layer of the retina, through connections with the suprachiasmatic nucleus and the pineal gland, switch the circadian rhythm from daytime to nighttime. The phylogenesis of the circadian rhythm was reflected in the dependence of the duration of the nighttime sleep of mammals on the diameter of the eyeball and the mass of the pineal gland. The activity of all the nerves of the eyeball led to the division of the nocturnal brain metabolism into NREM and REM phases. These phases correspond to two modes of the glymphatic system - electrochemical and dynamic. The first is responsible for the relaxation processes of synaptic plasticity and chemical neutralization of toxins with the participation of water and melatonin. Rapid eye movement and an increase in cerebral blood flow in the second mode increase water exchange in the parenchyma and flush out toxins into the venous system. Electrophysics of clearance and conductivity of ionic and water channels of membranes of blood vessels and astrocytes modulate oscillations of polarization potentials of water dipole domains in parietal plasma layers of arterioles and capillaries.

## 1. Introduction

### 1.1. The uniqueness of water physics

The relationship between the physiologies of the heart and brain is manifested in the transfer of pathologies between the nervous and cardiovascular systems [1-5], and in the reactions of the heart rhythm to changes in the human psyche [6, 7]. The energy and signal symbiosis of the heart and brain is realized through the neurohumoral regulation of their electrophysiology [8–12]. The electrical interconnection between cortical neurons and their zonal blocks is carried out by dendrites, intercalary neurons, and associative connections that commute in the thalamus and other structures of the subcortex [13]. At the level of the neuropil, the mechanism of neurovascular communication operates, in which the molecular dynamics of water ensures the localization of ion-mediator and trophic communications in areas with increased bioelectric activity [13-15]. Water as a key metabolite and the basis of brain fluid media determines the features of its energy at the cellular level, and anomalies in the thermodynamics of water and its solutions are responsible for the stabilization of the metabolism of the awake mammalian brain at a temperature of ~35 to ~39 °C [17–19]. In humans, normal brain temperature (T) during wakefulness is $T_b=36.9\pm0.4$ °C [18], while in sleep it decreases by ~0.5 and equals $T_S=36.5$ °C [24–27]. These values are close to $T_w=34.5$ °C, in the vicinity of which the isobaric heat capacity of water ($C_P$) has a minimum, and the dependence on T (TD) of the isochoric heat capacity has a weakly pronounced bend [20, 21]. It can be assumed that the features of the molecular physics of water in the vicinity of $T_w$ [21–23] are responsible for the mechanism of stabilization of normal brain T in humans during wakefulness and sleep, and play a key role in the thermodynamics of the glymphatic system of the brain.

### 1.2. Electrophysics of the brain and blood

The main elements of the body's electrical network are cardiomyocytes of the heart and blood vessels, as well as neuronal synapses and cerebrospinal fluid (CSF) of the cerebral cortex. Ionic currents in gap junctions of working cardiomyocytes and in channels of synapse membranes generate action potentials (AP) and electromagnetic waves (EMV). AP ensure the functioning of the body's internal communication systems responsible for its energy, dynamics and somatosensory. The maximum propagation velocity of AP is achieved in myelinated fibers (~100 m/s). According to the laws of electrical and electromagnetic induction, EMVs polarize blood plasma and intercellular fluid (ISF) of the cerebral cortex, and also excite magnetic vortices in neurons and in oscillatory circuits of neural networks [28, 29]. The limiting velocity of EMV propagation (C*) through the blood vessels and aqueous media of the brain is equal to the speed of light divided by the refractive index of water (n~1.3) [6, 28-30]. With such a speed, the potentials of the electric and magnetic fields of the heart and brain propagate through the liquid media of the body. The amplitude-frequency spectra of potentials are measured at certain points of the body and head in the form of electrocardiograms (ECG), electroencephalograms (EEG) and magnetoencephalograms [6, 28, 30-35]. During operations on the brain, it is possible to register potentials on the open surface of the cortex in the form of electrocorticograms (ECoG).

Molecular dynamics and charge separation in the brain is driven by the energy of glucose oxidation and electrolyte hydration. In contrast to the brain, the energy of the heart is 60-70% determined by the metabolism of fatty acids that are more energy-intensive than glucose, and therefore the specific power of the heart is ~2 times greater than the specific power of the brain [32]. In addition, due to the small proportion of current synapse dipoles oriented orthogonally to the scalp surface, the amplitudes of the EEG spectrum are of the order of microvolts (Fig. 1) at frequencies from Hz to kHz. Organization of cardiomyocytes in myocardial syncytia and synchronization of their current dipoles during the cardiocycle ensure their integration into the current macrodipole of the heart, the field of which has potentials of the order of millivolts, and frequencies from ~0.1 Hz to ~50 Hz (Fig. 1). The spatiotemporal distribution of the potentials of this field reflects the dynamics of the macrodipole vector of the heart [33, 34] and is recorded in standard leads with ECG potentials.

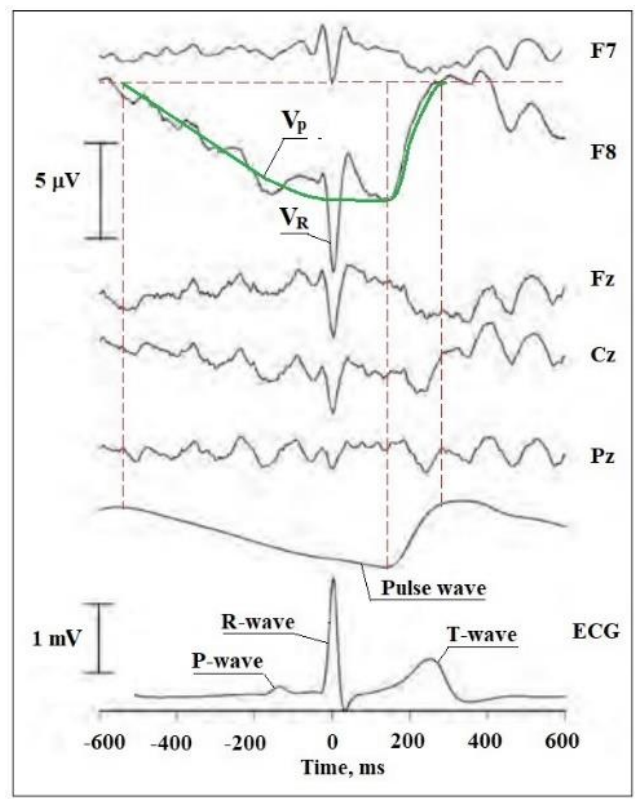

**Fig. 1.** Synchronous ECG and EEG spectra. P-, R- and T-wave are the potentials of the cardiogram teeth, Pulse wave is the potential of the blood pulse wave; F7, F8, Fz, Cz, Pz – points of EEG potentials with background alpha-rhythm; $V_R$ and $V_{pw}$ are the EEG potentials corresponding to the R-wave of the cardiocycle and the pulse wave in the blood plasma. Figure adapted from [6].

Gap junctions provide movement of the depolarization front of myocardial cardiomyocytes at a speed of ~1 m/s. The dielectric relaxation time of water ($\tau_D$), which characterizes the dynamics of the collective reorientation of molecules at 37°C, is ~7 ps [22, 36, 37], which ensures the propagation of EMV through the liquid media of the body and brain at a rate of C*.

The blood contains formed elements (mainly erythrocytes), their volume fraction (hematocrit) is normally ~40%. The rest of the volume falls on the plasma, which is an aqueous solution of electrolytes (2-3%) and proteins (up to 7%). CSF and ISF have approximately the same composition. The water content in the parenchyma of the cerebral cortex reaches 84%, and the CSF, which consists of ~99% water, normally occupies ~10% of the intracranial volume. From the point of view of electrophysics, physiological fluids are electrolyte solutions with high electrical conductivity ($\gamma$). For example, the $\gamma$ values of CSF, plasma, whole blood and muscle tissue at 37 °C are (in S/m) 1.8; 1.6; 0.54 and 0.66 [38], respectively. For comparison, $\gamma$ of an isotonic solution (0.9% NaCl) is 0.03 S/cm, and that of chemically pure water is 5.5 μS/cm [38]. However, the dipole moment ($\mu$) of a water molecule increases from 1.8 D to ~2.8 D during the transition from the gas phase to the liquid phase [20] due to the spontaneous self-organization of water hydrogen bond dipoles (HBs) into clusters and domains [20, 22, 23, 36].

The electrical continuity of the extracellular space of the parenchyma in ion exchange disorders of various etiologies allows the propagation of the depolarization front (SD) of neurons through the gray matter of the brain of humans and animals [39]. The SD rate is limited by the diffusion of ions ($Ca^{2+}$, $Na^+$) in the ISF and varies in different areas of the brain within 0.5–10 mm/min [40, 41]. On the other hand, from the time of understanding the meaning of a word by a person, 100-150 ms [42], it follows that the velocity of propagation of electrical signals between neuron systems along a network of chemical synapses is estimated at 0.1-1 m/s. There are also cases of a significant increase in the volume of CSF while maintaining the capacity of the brain. Moreover, a person with hypertrophy of the fourth ventricle and cisterns of the occipital part of the brain developed a phenomenal memory [43].

**1.3. Relationship between EEG amplitudes and frequencies**

The mechanism of neurovascular communication ensures an increase in blood flow to active areas of the cortex [11, 12, 44, 45]. The corresponding relationship between the electrophysics of the brain and the heart, in principle, should be manifested at the level of the EEG and ECG [6, 30]. The EEG frequency and amplitude spectra mainly reflect the dynamics of the scalp distribution of potentials induced by currents in the postsynaptic membranes of cerebral cortex synapses [32, 46, 47]. The chemical synapse and gap junction are modeled by a current dipole $P_j(t)$ (Fig. 2):

$$P_j(t) = Jd = \dot{q}(t)d . \qquad (1.1)$$

The recharging current $J$ in (1.1) changes reversibly from zero to a maximum in time $\tau$ and the reciprocal value $1/\tau$ will correspond to the frequency ($\nu$) of the electric field potential associated with the current dipole. The value of $\tau$ is of the order of $\sim 10^{-2}$ s and, accordingly, the oscillation frequency of the dipole field is of the order of 100 Hz.

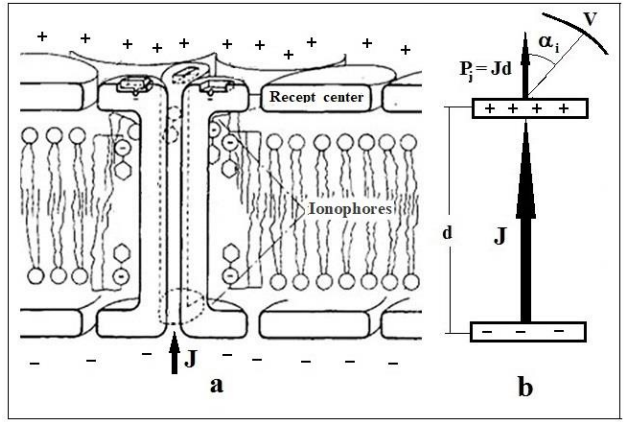

**Fig. 2.** Scheme of the site of the postsynaptic membrane of the inhibitory synapse – **a)**, (+) and (–) denote ions $K^+$, $Na^+$, $Cl^-$; **b)** Model of the current dipole ($P_j$) of the synapse ($J$ is the ion current, $d$ is the membrane thickness, $\alpha_i$ is the angle between the vector $P_j$ and the direction to the EEG potential $V$ tapping point on the scalp. The figure is adapted from [30].

The potential of the synapse current dipole ($\varphi_i$) at a distance r is given by the formula [48]:

$$\varphi_i \sim \frac{\cos\alpha_i}{\varepsilon\gamma} \frac{P_j}{r^2} , \qquad (1.2)$$

$\varepsilon$ is the dielectric constant of the parenchyma of the cerebral cortex, equal to 85 [48] and close to $\varepsilon$ of water; $\gamma$ is the electrical conductivity of the membrane. The value of $\varphi_i$ will be close to zero in the membrane itself and will be maximum in the current direction (Fig. 2). Potential $V$ in the amplitude spectrum of EEG can be expressed as the sum of projections $\varphi_i$ from all synapses located in the cylindrical cortical column under the $V$ pickup point on the scalp (Fig. 2b). The amplitude and sign of $V$ are determined mainly by the level of synchronization of the activity of excitatory or inhibitory synapses, the current dipoles of which are correlated in space.

Assuming the value of $V$ to be proportional to the potential difference across the membrane, the product $Vq$ can be related to the energy of the current dipole, and the expression $qV/\tau$, to its power. The sum over all i-synapses of the active cortical column will give an adequate measure of the total power of the coherent ensemble of synapses. In this case, the EEG power spectrum can be expressed by the formula:

$$\Psi_{EEG} \sim \sum (V\nu)_i.$$

The value of $\Psi_{EEG}$ will depend not only on the angle $\alpha_i$, but also on the anatomical and trophic features of the cortical zones due to their functional specification [47, 49], which reflects the influence of sensory electrophysics on the genesis of the neocortex [15, 30]. However, taking into account the isotropic distribution of the density of capillaries [50] and current dipoles in the "synaptic brain" [51-53] over the cortex, it can be assumed that the specific power of the electrical activity of the cortex, and hence $\Psi_{EEG}$, have similar values throughout the scalp. This confirms the closeness of $(V\nu)$ values at all standard points of the scalp in the synchronous frequency and amplitude EEG spectra [57], as well as the observation that during sleep, high $\nu$ and low $V$ (~10 Hz, ~0.01 mV) pass into high $V$ and slow NREM sleep waves (0.5-4 Hz, 0.2 mV) [54, 55]. This implies a qualitative dependence for the EEG spectra:

$$V \sim const/\nu,$$

which is also valid for frequency and amplitude ECG spectra [57].

It is known [32] that sensory receptors transmit information to the brain by varying the repetition rate of spikes from action potentials (AP). In the process of development of the sensory systems of the brain of mammals, external signals of an electromagnetic nature and chemical factors led to an expansion of the frequency range of the electrophysiology of the brain. If the lower level of EEG frequencies ~0.01–1.0 Hz corresponded to the modulation of brain electrophysics by the rhythm of respiration and heartbeat [56], then from above the EEG frequency, the period of activity of current dipoles in the membranes of neurons and synapses was limited to 1–2 ms (~1 kHz).

With this in mind, when analyzing the electrical activity of the brain in the amplitude $V$ spectrum of EEG, characteristic frequencies of $V$ repetition are usually detected and correlated with known brain states. Within the physiological boundaries of frequencies $V$, the EEG amplitude spectra are conditionally divided into frequency ranges: delta (0.5-4 Hz), theta (4-8 Hz), alpha (8-13 Hz), beta (13-30 Hz) and gamma (30-100 Hz). As a rule, certain functions and states of the brain and heart are correlated with these ranges, so the chronometry of ECG and EEG frequencies is useful for clarifying the nature of brain activity. In [30, 57], for example, from the analysis of changes in the ECG and EEG frequency spectra at standard potential tapping points, regularities were revealed in the reactions of the heart and brain to vocal acoustics, as well as to the activation of the visual system by light of different wavelengths and heat.

### 1.4. Polarization of parietal layers of blood plasma

The autonomy of the heart metabolism and the mechanism of spontaneous periodic depolarization of the membranes of the sinoatrial node (automaticity) correspond to an open self-oscillating system, the rhythm of which can play the role of a basic pacemaker of electrophysical processes in the brain [6, 30].

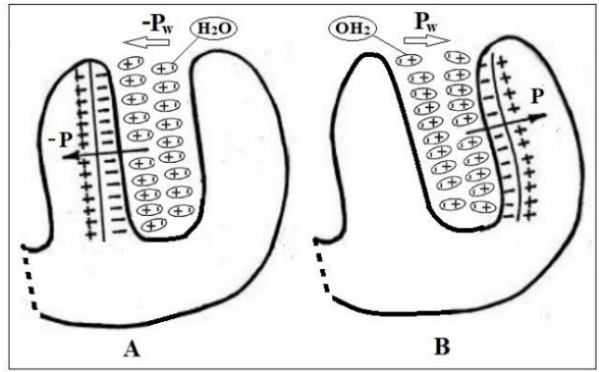

**Fig. 3.** Scheme of depolarization of cardiomyocytes in the left ventricle. The values of the *V* ECG are recorded in the chest leads. **A**) - depolarization of the left half of the interventricular septum (P-wave). **B**) - depolarization of the left ventricle (R-wave). $P_w$ – layer polarizability potential; ***P*** is the dipole moment of water domains in near-wall plasma layers. Figure adapted from [33, 34].

The heart, as a multipolar current and magnetic dipole, during the cardiocycle induces electric and magnetic fields that are variable in time and space, which propagate at a speed of C* throughout the entire circulatory system of the body. Extreme ***P*** values of local syncytia from cardiomyocytes occur on the inner walls of the left ventricle during depolarization (Fig. 3) [33, 34]. Under the influence of these fields, water molecules in the parietal plasma layer in the left ventricle self-organize into dynamic domains and supramolecular structures of HBs with high values of μ and the polarizability potential of the layer ($P_w$) orthogonal to the ventricular wall (Figure 3). The value of $P_w$ will be proportional to the product of the surface charge density of the layer ($q_w$) and its thickness (d);

$$P_w \sim q_w d. \quad (1.3)$$

When blood is pushed into the aorta, the $P_w$ wave propagates along the parietal layer of the plasma of the arteries and capillaries of the brain as along an equipotential surface. This is facilitated by immobile blood in the parietal (lubricating) plasma layer free from formed elements, the thickness of which is proportional to the diameter (*d*) of the blood vessel. For example, at a blood flow velocity of 0.2–0.9 mm/s, the plasma layer in capillaries $d \sim 7$–12 µm ranges from 0.4 to 1.6–2 µm, while in vessels with $d$ about 500 µm, its thickness reaches 15–45 µm [58]. Due to the dielectric properties of this plasma layer, blood vessels can be modeled as cylindrical waveguides, along which transverse electric and magnetic waves propagate, corresponding to displacement currents in the plasma layer [59]. The biomagnetic signals of the heart are two orders of magnitude stronger than the biomagnetism of the brain and, in principle, make it possible to diagnose coronary heart disease [60]. The technogenic magnetic field is 3, and the terrestrial one 6 orders of magnitude stronger than the magnetic field of the heart, so its influence on the electrophysics of the brain is practically impossible to detect, in contrast to the effects of $P_w$ [61].

The polarization of clusters in the HBs network in the layer will be promoted by electrolyte hydrates, molecules with high µ, and the negative charge of the glycocalyx of endothelial cells bordering the plasma [62, 63]. Note that the potential difference $P_w$ between the left ventricle and capillaries of the cerebral cortex corresponds to the distribution of ECG potentials inside the body. For example, this difference between the cerebral cortex and jugular blood flow in animals is of the order of magnitude of 1–5 mV characteristic of ECG [8, 35].

### 1.5. Electrophysics of the parenchyma of the cortex

Potentials $P_w$ of the near-wall layer of arteries and capillaries generate an electric field ($\varphi_w$) in the walls of vessels and outside them, similarly to the current dipoles of synapses $\varphi_i$ (1.2) [47, 63]. The value of $\varphi_w$ at a distance $r$ from the outer surface of the plasma layer, taking into account expression (1.3), can be expressed by the formula:

$$\varphi_w \sim \frac{\cos\beta}{\varepsilon} \frac{P_w}{r^2}, \qquad (1.4)$$

$\beta$ is the angle between the vector $\boldsymbol{r}$ and the perpendicular to the axis of the vessel. Thus, the sum of the vectors $\varphi_w$ and $\varphi_i$ will determine the magnitude and direction of the field strength ($V_{ex}$) at each point of the parenchyma. The $V_{ex}$ projection on the scalp is fixed by the EEG amplitude spectrum, in which $\varphi_w$ from the superficial arteries manifests itself as a $V_R$ wave (Fig. 1).

The maximum value of $\varphi_w$ in the parenchyma of the cortex and $V_R$ on the scalp will be generated by vessels whose axes are parallel and the $P_w$ vectors are orthogonal to the surface of the cortex. The $V$ values between the retina and cornea (0.4–1.0 mV) [64] and the difference in $V$ values between typical EEG and ECoG [33] indicate that the skull bone, due to low $\gamma$ [38], weakens $V_R$ by almost an order of magnitude. With this in mind, from Fig. 1 follows the estimate of $\varphi_w$ for the superficial artery ~50 µV. Taking the value of $d$ in (1.3) proportional to the radius of the vessel, for typical parameters of an artery (~1.5 mm) and a capillary (~6 μm) [50], we obtain from (1.4) for capillaries $\varphi_w$~0.2 μV. This value turns out to be of the same order of magnitude with the calculated $\varphi_i$ of the current dipole of the synapse 0.1 μV [6]. The density of capillaries in the cerebral cortex is 600-800/mm$^3$ [65], and that of synapses is $5·10^8$/mm$^3$ [51], therefore, on the scalp, total $V$ from synapse ensembles correlated in space and time are fixed, in contrast to $V$ generated by $P_w$ in capillary ensembles with small values of the angle β in (1.4).

In [6], $V_R$ and $V_P$ waves in the EEG spectrum were detected using a special program for frequency filtering of the EEG background alpha rhythm and by 450 EEG signal averaging (Fig. 1). Moreover, the $V_P$ pulse wave appeared only in lead F8, and it is absent at other points of the scalp, including point F7 on the left side of the head. This result can be explained by the exact location of the F8 point above the superficial frontotemporal artery and the violation of such proximity in the case of the F7 point and the left analogue of the frontotemporal artery. Under the points FZ, CZ and PZ is the superior sagittal sinus of the venous system, devoid of pulse waves. Note that the results of [6] are consistent with the data of [66], in which $V_R$ waves were detected in the frontal region and against the background of $V_P$ waves with a duration of 300–600 ms. In [67], on changes in intracranial pressure associated with the arterial pulse wave, it was found that the EEG $V_P$ wave follows the ECG *R*-wave with a delay of 40–160 ms.

The electrophysics of myelinated neurons of the neocortex is practically not involved in the formation of $V_{ex}$. The metabolic energy of neurons is spent mainly on synaptic activation and the biochemistry of synaptic plasticity [44, 68]. In this case, the entire energy of ionic currents in the nodes of Ranvier is spent on the generation of magnetic vortices in the saltatory mechanism of AP [28]. Thus, the potentials of current dipoles of synapses ($\varphi_i$) and $P_w$ of capillaries ($\varphi_w$) are responsible for the formation of the extracellular field $V_{ex}$ in the cortical parenchyma. Due to the isotropic distribution of capillaries in the cortex, the amplitude and frequency spectra $\varphi_w$ at rest with eyes closed will be approximately the same everywhere, and in the spectra $\varphi_i$ there will be areas with increased alpha activity [69] (Fig. 1). During mental work, the neurovascular connection should manifest itself by synchronization of changes in the frequency-amplitude spectra $\varphi_i$ and $\varphi w$ in areas of the cortex with increased neuronal activity.

The $P_w$ values in the blood vessels oscillate synchronously with the frequencies of the cardiocycle waves. In addition, within each cardiocycle, the polarization of the parietal plasma layer of arteries and arterioles is perturbed by a pulse wave caused by elastic deformation of smooth muscles. In the capillary system, the pulse wave disappears and only oscillations of the amplitude $P_w$ remain, synchronous with the waves of the cardiocycle. In the EEG spectrum, the pulse wave corresponds to the *V*-wave ($V_P$ in Fig. 1), which is modulated by the waves of the QRS complex ($V_R$ in Fig. 1) and the T-wave of the cardiocycle. Similarly, in all EEG leads, in principle, a weak *V*-wave can appear, corresponding to the *P*-wave of the cardiocycle. This wave represents a wave of depolarization of cardiomyocytes of the right atrium and corresponds to $P_w$ in the superficial veins and superior sagittal sinus.

The distribution of $V_{ex}$ at a rate of C* in the cortical parenchyma can occur through the extracellular space of the parenchyma, which is convoluted tunnels and channels 38–64 nm wide [70, 71]. ISF is saturated with negatively charged hyaluronic acid and proteins that effectively bind water and $Ca^{2+}$, $Na^+$, and $K^+$ cations [72, 73]. In this case, ISF loses bulk fluidity [71] and the diffusion of water, ions, and molecules decreases by a factor of 3–5 compared to diffusion in free water [73, 74]. The structure and dynamics of ISF, by blocking the leakage of $Ca^{2+}$ and neurotransmitters from synapse clefts [74], contributes to the propagation of white noise in resonant mechanisms of information transmission and detection in neural networks [28, 44, 75-79].

On the other hand, the high efficiency of synapses is ensured by the small width of synaptic clefts (10–20 nm) at d~1–2 μm [81] and the acceleration of diffusion of the $Ca^{2+}$ cation and polar mediators γ-aminobutyric acid (GABA), glycine, and glutamate in local fields $\varphi_i$ of neuron receptors and astrocyte transporters. In the case of mediators, the field effect enhances the charge that mediators acquire by forming complexes with ions present in the gap [81, 82]. For example, GABA, upon leaving the vesicle, binds to two $Na^+$ cations and one $Cl^-$ anion. Thus, the activity time of GABA, glycine, and glutamate synapses is about 100-200 ms, which determines the speed of communication in the neural systems of cognitive functionals and is manifested in the EEG spectra at frequencies of ~5-10 Hz. It should also be noted that the preservation of GABA in water in its linear conformation with the maximum value of μ [83-85] will increase $P_w$ and its polarization effects on the blood-brain barrier and on ISF dynamics around the brain capillaries.

In this work, in order to elucidate the mechanisms of participation of water in the thermodynamics and electrophysics of the human heart and brain, a comparative analysis of the activation energies of the temperature dependences of the electrophysical and dynamic properties of water and physiological fluids of humans and animals was carried out.

## 2. Methods and materials

To establish the role of water in the thermodynamics and electrophysics of cerebrospinal fluid, plasma, and blood, a comparative analysis of the temperature dependences (TDs) of their dynamic and structural parameters in the T range from ~25°C to ~50°C was performed using bimodal Arrhenius approximations ($F_A$) [23, 86]:

$$F_A = T^\beta \exp(\pm E_R/RT) = \exp[(\pm E_T \pm E_R)/RT] = \exp(\pm E_A/RT). \quad (2.1)$$

R is the gas constant (8.31 J·mol$^{-1}$·K$^{-1}$). According to (2.1), the activation energy ($E_A$) or the thermal effect of the rearrangement reaction of the molecular or supramolecular structure of a liquid is the algebraic sum of the thermal component ($E_T$) and the electric component, including the energy of Coulomb ($E_R$) and van der Waals (vdW) interactions [86, 87]. In general, most TDs can be divided into T-intervals, in which (3.1) will give the $E_A$ value corresponding to the dominant mechanism of water molecular dynamics. Note that taking into account vdW under the assumption of weak HBs made it possible to calculate the position of the maximum TD of water density with an accuracy of ~2% [87]. The value of β was chosen taking into account the physical nature of the structural or dynamic parameter and the role of thermal energy in the reaction mechanism limiting TD. These parameters may play a role in the electrophysics of water and physiological fluids (FFs).

The values of $E_T$, $E_R$, $E_A$ and β for the main parameters of water – the diffusion coefficient ($D_w$), dynamic viscosity (η), $τ_D$, ε, isobaric heat capacity ($C_P$), fluctuation amplitude of the angle HB (δ), tetrahedral index HBs ($q$), are shown in Table 1. The method of applying $F_A$ approximations to determine $E_R$ is shown using the TD $C_P$ of water at 760 mmHg as an example in the range of 29-40 °C (Fig. 4).

For γ and pH, by analogy with $D_w$ and η, the values of β were taken to be 1 and 0, respectively. At β=0, $E_T$=0 and $E_A$=$E_R$ were considered. The $E_R$ estimate for γ water and FFs in the range of 0-50 °C was obtained by subtracting the average in the T-interval $E_T$~2.6 kJ/mol from the $E_A$ modulus (Table 2). The specific electrical conductivity of water ($γ_w$) depends on the following parameters [88]:

$$γ_w \sim K_w^{1/2}(γ_H^+ + γ_{OH}^-),$$

$K_w$ is the dissociation constant of water into proton and hydroxyl, $γ_H$ and $γ_{OH}$ are specific conductivities.

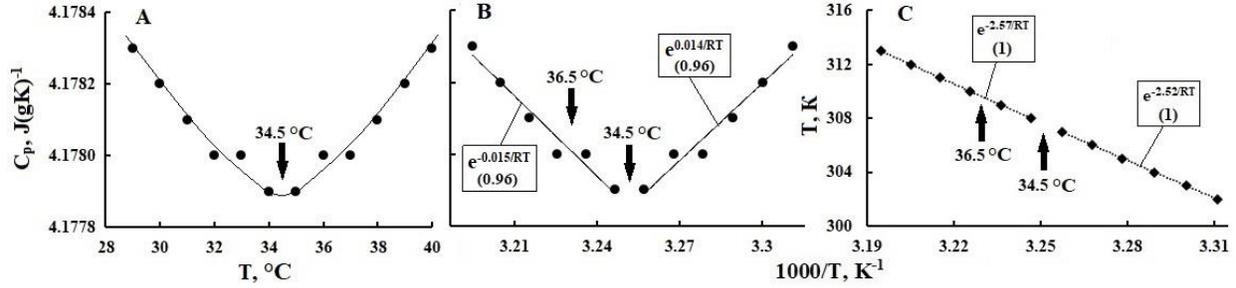

**Fig. 4.** Dependences of specific heat capacity of water ($C_P$) on T in °C, the line is the envelope (**A**). Dependences on the inverse $1/T$ in $K^{-1}$ $C_P$ (**B**) and T (**C**); lines are $F_A$-approximations. Const in exponents - $E_A$ (kJ/mol), in parentheses - $R^2$. Modified from [23].

The calculated and empirical TDs of these parameters in the range of 0–100 °C are presented in [88] in the form of tables. The intervals in TD were equal to 5°C, so the $E_A$ estimates in the ranges of 5-25°C and 5-35°C were almost the same. TDs of $K_w$ and pH = -lg[$H^+$] correlate and for their approximation instead of (2.1) a linear function of the form was used:

$$F_A = \text{const } T^{-1}, \quad \text{const} = \frac{E_A}{R\,ln10}. \tag{2.2}$$

Note that the pH of pure water characterizes the Coulomb interactions inside the cell, which determine the probability of $H^+$ exit from the cell and the equilibrium dissociation constant of water. The same interactions of the central water molecule with the nearest environment determine the dielectric relaxation time and the friction force, which is proportional to the dynamic viscosity [22, 86].

**Table 1.** Values of β for approximations (2.1) and activation energy (in kJ/mol) for water characteristics in selected temperature ranges (see text).

| Parameter | β | ΔT (°C) | $E_T$ | $E_R$ | $E_A = E_T + E_R$ | Ref. |
|---|---|---|---|---|---|---|
| $D_w$ | 1 | 30-50 | -2.6 | -13.6 | -16.2 | [86] |
| η | 0 | 26-50 | 0 | 14 | 14 | [86] |
| $\tau_D$ | -1 | 30-60 | 2.6 | 14 | 16.6 | [23, 86] |
| $C_P$ | 1 | 29-34 | -2.52 | 2.53 | 0.014 | Fig.4 |
| | | 34.5 | -2.55 | 2.55 | 0 | [23] |
| | | 36-40 | -2.57 | 2.56 | -0.015 | |
| q | 0 | 28-50 | 0 | 2.7 | 2.7 | [30] |
| δ | 1 | 13-60 | -2.6 | -0.1 | -2.7 | [86, 87] |
| ε | | 0-25 | 2.4 | 0.6 | 3.0 | [22, 23, 86] |
| | | 30-45 | 2.6 | 2.6 | 3.7 | |
| $n^2$ | -1 | 25-36 | 2.55 | -2.42 | 0.13 | [23] |
| | | 37-47 | 2.55 | -2.37 | 0.18 | |

The accuracy and reliability of TDs measurements of dynamic and electrophysical parameters of blood, plasma, and CSF depend on the degree of adequacy of the composition and concentrations of models and samples of FFs, as well as the conditions of experiments in vitro and in vivo. For example, due to the lack of reliable pointwise TDs of blood and plasma pH [90, 91], linear extrapolations of pH values at ~20°C and 38°C, obtained as early as 1948 [90], still appear in the literature. They look like:

$$pH_t = pH_{38} + const\ (38 - t), \qquad (2.3)$$

where t is T in °C, $pH_{38}$ is ~7.4 for blood and plasma, and *const* for them is 0.0147 and 0.0118, respectively [90]. From the analysis of TD of disparate reference pH values of pure water in the range from 10 to 50 °C (Fig. 5A), a break in $F_A$ at ~25 °C (Fig. 5B) follows, which is absent in extrapolations (2.3) (Fig. 5C).

In this work, a comparative analysis of the known TDs dynamic and structural parameters of FFs, measured by the methods of dynamic light scattering (DLS) and circular dichroism (CD), was carried out. From DLS, a conditional hydrodynamic radius ($R_H$) is obtained, appearing in the Stokes-Einstein equation $D \sim T/(\eta R_H)$. The DLS method makes it possible to evaluate the dependence of the mobility of dissolved substances on their structure through changes in the dynamics of the medium. The CD method, by measuring the angle of ellipticity ($\theta$) of protein solutions, makes it possible to determine changes in the proportion and configuration of chiral fragments of alpha helices in them.

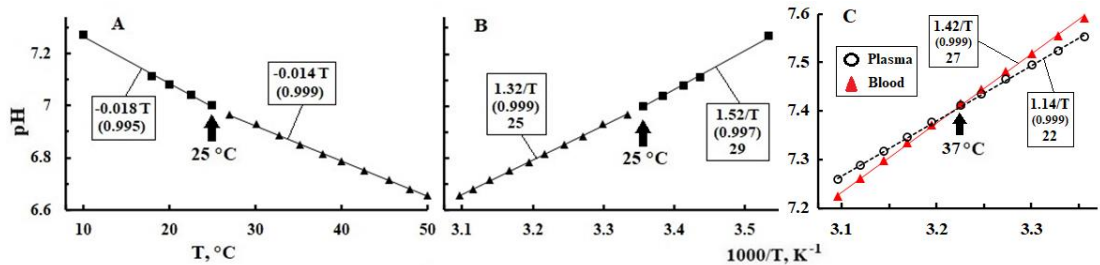

**Fig. 5. A**. Dependence of pH of pure water on T °C (points) and its linear approximations (lines). **B**. The dependence of pH water on the 1/T (points) and its approximation by function (2.2) – lines. Boxes show $E_A$ values in kJ/mol. **C**. Dependences on 1/T blood and plasma pH (points) and their linear approximations by function (2.3). Initial pH TDs are taken from [90] and handbooks.

The method of neutron scattering determines the root-mean-square displacement ($u^2$) of atoms in alpha-helices, which makes it possible to calculate their elasticity and the level of interconnection of proteins with their hydration shells (HS). Interactions of proteins with water can distort the Arrhenius shape of TDs, moreover, the reliability of TDs depends on the degree of adequacy of the correspondence between the measured values and the FFs parameters. Therefore, in analyzes of the known TDs θ, $R_H$ and $u^2$ of various model solutions, the main attempt was made to identify qualitative correlations between $E_A$ values. For θ and $R_H$, β=0 was taken.

We also analyzed the known TDs of the following water parameters and FFs:
- ellipticity (θ) of model solutions of hemoglobin (Hb) and other blood proteins;
- optical activity ([α]) of saccharides in saline solutions (0.9% NaCl), according to the angle of rotation in degrees (φ) of polarized light, taking into account the dependence of φ on the concentration of the substance and the specific angle of rotation, the tabular value of which is determined for light with λ=589 nm ([α]$_D$);
- γ of water and NaCl and KCl solutions, as well as spin-lattice relaxation times ($T^1$) of NaCl solutions ($T^1_{NaCl}$) and sea water ($T^1_{Sea}$);
- solubility of oxygen gases ($α_{og}$) and carbon dioxide ($α_{cd}$) in water, isotonic solution (0.9% NaCl) and plasma;
- γ and pH of water and human blood at different values of hematocrit (Ht20%, Ht40%, Ht60%).

The accuracy of data on TDs of water parameters and FFs allows rounding the received $E_A$ to integer values without losing the ability to identify trends in their changes. The $E_A$ values in kJ/mol are given in the Table 2 or in the graphs in terms of $e^{\pm const}/RT$, where const = $E_A$. Empirical data for TDs were imported from published sources. References to these materials are given in figure captions and in Table 2. Graphs were digitalized with the use of Paint computer application, when necessary. MS Excel application was used to plot TDs and their approximations. The extent of proximity of value $R^2$ to 1 on T-intervals was chosen as the reliability criterion, for approximations, in various T-intervals. The values of extreme T were marked on the graphs by arrows.

### 3. Results

Plots of known TDs and their approximations are shown in Fig. 5-10, and the values of the T and $E_A$ ranges are given in Table 2. Regarding η and D of CSF, it is known [92, 93] that CSF, like plasma, is a Newtonian fluid and its dynamic viscosity is normally not affected by proteins and cells. At 37 °C, the η value of CSF lies in the range of 0.7–1 mPa s, while the η values of water and blood plasma are ~0.7 and 1.5 mPa s [94], respectively. Hence it follows that the values of $E_A$ for the dynamic characteristics of water, liquor and plasma should be close. The $E_A$ estimate for CSF γ was obtained using γ at 25 °C (1.45 S/m) and two γ values at body temperature 37 °C obtained by direct current measurement – 1.79 S/m [95] and the average γ from 16 measurements using the Magnetic Resonance EIT – 1.89 S/m [38]. The calculation gave $E_A$ 12.6 kJ/mol and 17.5 kJ/mol, with an average $E_A$ of ~15 kJ/mol. These $E_A$ values are shown in Table 2.

**Table 2.** Activation energies ($E_A$) of temperature dependences of viscosity (η), conductivity (γ, S/cm), pH, spin-lattice relaxation time ($T^1$) for water, electrolyte solutions, plasma, CSF, blood and hydrodynamic radius ($R_H$) of the model solution Hb human.

| Fluid | | Parameter | ΔT (°C) | $E_A$ (kJ/mol) | Reference, Fig. N |
|---|---|---|---|---|---|
| Blood | Ht 20% | γ | 30-35 36-40 | -32 -21 | [96, 97] |
| | Ht 40% | | | -36 -23 | |
| | Ht 60% | | | -31 -20 | |
| | Ht 40% | pH | 20-50 | 27 | [90] |
| Plasma + Hb | | $R_H$ | 21-36 37-51 | -4 -16 | [19] |
| Plasma | | γ | 30-35 36-40 | -17 -11 | [96] |
| | | pH | 20-50 | 22 | [90] |
| | | η | 15-45 | 16 | [94] |
| CSF | | γ | 25-37 | -15 | [95] |
| Water | | $γ_w$ | 5-35 40-60 | -40 -33 | [88] |
| | | $γ_H$ ($γ_{ОН}$) | | -10.8 (-13) -8 (-10) | |
| | | $К^{1/2}$ | | -29 -25 | |
| | | pH | 10-25 27-50 | 29 25 | Справочники |
| Water+NaCl | | $T^1$ | 0-25 25-75 | -20 -16 -19.2 -16.3 | [23] |
| Sea water | | | | -19.2 -16.3 | |
| Water+Na$^+$ Water+ KCl | | γ | 15-40 | -16 | [98] |

## 4. Discussion

### 4.1. Anomalies in the thermodynamics of water in CSF and plasma

#### 4.1.1. Stabilization of brain temperature

Optimization of the thermodynamics of the metabolism of internal organs in the range of 36–38 °C is provided by the mechanisms of thermoregulation under the control of the hypothalamus and with the participation of blood water as the main coolant and heat exchanger in the brain [20, 21, 24, 99]. These mechanisms begin to deteriorate at T<35 and >40°C and completely fail at T<33°C and >42°C with a possible fatal outcome. Stabilization of the optimal T of the brain is ensured by the features of the dynamics and structure of water formed as a result of a dynamic phase transition in the vicinity of $T_h$=25 °C [22, 23, 86, 100, 101]. In this transition, the ice-like metastable structure of water, which consists mainly of hexagonal clusters (IhW), is rearranged into a mixed structure (IW) of chain and ring clusters with less than 6 molecules [20–23, 100–104]. Accordingly, the index q of tetrahedral HBs in the IW structure becomes smaller than in IhW [22, 86, 105]. The features of the thermodynamics of the HBs structure in the IW range of ~42–46°C are still preserved in the mechanism of minimizing the isothermal compressibility of water [23, 101]. At T>46°C, after the completion of the decomposition of small clusters into dimers and free molecules, bulk water turns into a homogeneous liquid, devoid of the thermodynamic features of the HBs network [106, 107]. At T > 42°C, the decomposition of HBs in the bulk of water is associated with the rearrangement of HBs into HS proteins, which initiates changes in their conformations, aggregation, and denaturation [108, 109].

In the range of 33–42 °C, the equilibrium thermal energy (RT~2.6 kJ/mol) is half the energy of HB breaking (~5–6 kJ/mol) [101, 110, 111] and an order of magnitude less than the energy of atomic vibrations in the H2O molecule [20, 21]. Note that the energy of coherent vibrations of 10 protons in a tetrahedral chain of 5 water molecules is 2.6 kJ/mol [100]. Therefore, the molecular dynamics in IW is limited by librations of free molecules and chains, as well as by fluctuations in the HBs network with $E_A^\delta$~2.7 kJ/mol. Cooperative librations of molecules and continual transformations of the HBs network [20, 110–112, 118] with $E_A$ of the order of energy vdW play a key role in the mechanism of anomalies in water thermodynamics in the range of 33–42 °C and, in particular, are manifested by a $C_P$ minimum in the vicinity of $T_w$=34.5 °C at normal pressure (Fig. 4, Table 1). For IW structures, the average number of tetrahedral HBs per molecule is ~3–3.5 [21]; therefore, an energy of ~15–18 kJ/mol is required for its release from the HBs cage.

The $E_A$ values for $D_w$, $\eta$, $\tau_D$ (Table 1), as well as for $T^1$, $T^1_{NaCl}$, $T^1_{Sea}$ (Table 2) are consistent with this estimate. Features of the thermodynamics of cooperative processes in HBs IW in the vicinity of $T_w$ do not manifest themselves noticeably on the TDs of the dynamic parameters of pure water (Table 1), but can affect the kinetics of electrophysical and biochemical processes in blood and CSF (Table 2).

Stabilization of water thermodynamics in the vicinity of $T_w$ is ensured by a fast mechanism of thermal energy dissipation in the bulk HBs network [113], consistent with HBs fluctuations and $H^+$ jumps within the cell [23, 86, 100, 114-118]. This mechanism corresponds to the closeness of the modules of the $E_T$ and $E_R$ values, which determine the kinetics of the IW structure rearrangement reactions at the level of conformations of the HBs supramolecular network [20, 100]. The synergism of $E_T$ and $E_R$, for example, appears on TD $\varepsilon$ at T in the range of 35-45 °C in the ratios $RT_w \sim E_T \sim |E_A^\delta|$ and $E_T \sim E_R^\varepsilon$ (Table 1), which determine the mechanism of matching the thermal excitation of HBs fluctuations with the exothermic rearrangement of the domain structure IW. In this case, the sum $E_T + E_R^\varepsilon$ turns out to be close to the value of energy of HB breaking.

The energy released during $H^+$ hopping within the cell ($E_R$) can serve as an alternative to $E_T$ in the rearrangement of tetrahedral configurations in the HBs network of the IW structure. This is confirmed by the following ratios of the modules $E_A$, $E_T$ and $E_R$ for $\delta$ and $q$: $E_T^q = 0$, $|E_R^\delta| \sim 0.1$ и $|E_A^\delta| = E_R^q = E_A^q$ (Table 1) [86, 87]. This consistency in the energies of rearrangement of the supramolecular structure of HBs apparently plays a key role in its stabilization and deformation in the range of 32–42 °C. For example, for $C_P$ at $T=T_w=34.5$ °C $|E_T|=E_R$ and $E_A=0$, while at $T<T_w$, $E_A>0$, and at $T>T_w$ $E_A<0$ (Fig. 4, Table 1). It follows that at the minimum of $C_P$, isothermal rearrangements of HBs occur, and at $T<T_w$, the exothermic $H^+$ jump inside the cell dominates over librations ($|E_T|<E_R$) and its energy ensures an increase in $q$ and the degree of clustering IW. At $T>T_w$, endothermic librations dominate over the rearrangement of bonds inside the cell ($|E_T|>E_R$), while $\delta$ increases [111, 117] and the probability of the molecule leaving the cell, which leads to a decrease in the clustering level IW and an increase in $D_w$.

Minimum $C_P$ in the vicinity of $T_w=34.5$ °C at 760 mm Hg shifts towards high T with decreasing pressure [20, 23]. In addition, the freezing temperature of plasma and cerebrospinal fluid is ~0.5 °C higher than that of pure water and the level of normal intracranial pressure is 7-15 mm Hg significantly below atmospheric. Hence, it follows that the features of the thermodynamics of pure water in the range of 33–42°C may well be preserved in brain fluids and ensure the stabilization of its metabolism during wakefulness and sleep [18, 24]. In accordance with this, in the range of 35-37 °C, breaks in $F_A$ TDs of the following FFs parameters are detected: γ blood and plasma (Fig. 6A), $α_{og}$ plasma (Fig. 7A), φ, $[α]_D$, $R_H$ and θ of model solutions of sugars and proteins (Fig. 8, Fig. 10, Fig. 11, Fig. 12A, Fig. 13A). It should be noted that point-by-point measurements of these parameters are usually made under thermostationary conditions and on immobile FFs. The calculated arterial and plasma pH TDs intersect at $T_b$ at normal pH (~7.4) (Figure 5C).

The thickness of the subarachnoid layer of cerebrospinal fluid (LiA), bordering the cerebral cortex, is comparable to the thickness of the parenchyma of the cortex, and they are separated by a thin and porous soft shell, the thermal conductivity of which in the transverse direction is close to that of water [119, 120]. Under such conditions, LiA serves as a "thermostat" of the brain, stabilizing the metabolism of the cortical parenchyma at $T_b$ by removing and transferring excess heat to the arteries immersed in LiA and the venous system of the superior sagittal sinus [120]. In the capillary segment of the Virchow–Robin canals (KVR), the efficiency of heat transfer between ISF and blood is higher than in the arterial segment due to the absence of smooth muscles on the capillaries and the low velocity of arterial and venous blood in them [120]. Thus, the features of the thermodynamics of bulk water in the range of 33–42 °C can be extrapolated to the thermodynamics of ISF, taking into account the specific effect on the structure and dynamics of IW of electrolytes, $CO_2$, and potentials $φ_i$ and $φ_w$.

### 4.1.2. Dynamics of proton and electrolytes in water and blood

The molecular dynamics of water and FFs in the range of 33-42°C are determined mainly by the synergism of Brownian fluctuations and $H^+$ jumps inside and outside the cell. In a cell of the HBs tetrahedral network, the jump of $H^+$ to the vacant $sp^3$ oxygen orbital of its own or neighboring molecule is consistent with the rotation and displacement of the molecule. In the subsequent process of relaxation of the cell dipoles, the HB is switched and the cell is rearranged [23, 114]. The synergy of $H^+$ jumps with Brownian motion and dipole relaxation [23, 115] is responsible for the dynamics of liquid water at the level of HBs cells. In confirmation of this scheme, in the structure of IhW at $T<T_h$, the values of $E_A$ of the molecular shift and rupture of HB are equal to 7.8 kJ/mol [86] and are close to $E_A$=8.4 kJ/mol, which follows from the TD difference in the Raman spectrum, which characterizes the rearrangement of HBs in the range of 5-80 °C [110, 111]. In the vicinity of $T_h$, the lifetimes of the HB and the cell itself are compared [22, 86].

The dynamics of H⁺ in pure water IW depends on the value of $q$ and defects in the tetrahedral structure of the cells of the HBs network [105, 121], while in FFs the relaxation of perturbations in the IW structure is accelerated in the fields of electrolytes and their character changes in the hydration shells of proteins [108, 122].

For $\gamma_w$ and water pH, $E_R$ sums up the endothermic reaction of H⁺ release from the cell and the exothermic reaction of H⁺ and OH⁻ hydration. Therefore, $E_A$ for $\gamma_w$ is equal to the sum of $E_A$ for $\gamma_H$ and $K^{1/2}$ in the corresponding T-intervals (Table 2). Due to the low probability of H⁺ exit from the cell and the reversibility of the water dissociation reaction, the equilibrium concentrations of H⁺ and OH⁻ are equal to $10^{-7}$ mol/L. At this concentration, the average distance between H⁺ is ~0.25 μm, and $E_A$ for the pH of water and plasma is 25 and 22 kJ/mol. In an electric field, $E_A$ for $\gamma_H$ decreases by half (Table 2) due to an increase in the probability of H⁺ exit from the cell and subsequent tunnel hops along the HBs chains of water and protein alpha helices according to the Grottuss mechanism [123]. Since the contribution of γn to the conductivity of electrolyte solutions is negligible, the $E_A$ values for γ plasma and cerebrospinal fluid are close to the $E_A$ values for the dynamic characteristics of water and η plasma (Tables 1 and 2). Hence it follows that the kinetics of ion homeostasis in the cortical parenchyma is mainly limited by the molecular dynamics of water in ISF.

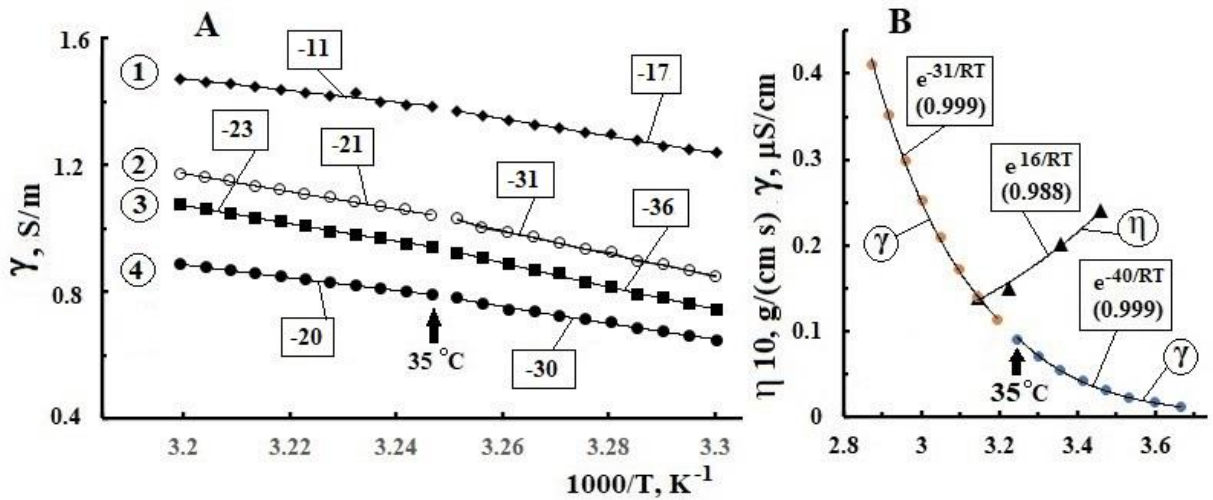

**Fig. 6.** Dependences on 1/T of the specific electrical conductivity (γ) and viscosity (η) of liquids (dots) and their $F_A$-approximation (lines). **A)** 1 - plasma, 2, 3, 4 - canned blood with hematocrit 20%, 40%, 60%, respectively (Within $E_A$ in kJ/mol, $R^2 > 0.99$ for all $F_A$). **B)** water (γ) and blood plasma (η). The initial data are taken from [94, 96, 97].

The average pH values of CSF, arterial and venous blood are normally 7.33, 7.39 and 7.31, respectively. The acidity of venous blood increases by ~1%, and the concentration of $H^+$ by ~1.2 times due to an increase in $H_2CO_3$ due to the transfer of $CO_2$ from the parenchyma to the capillary blood [91]. The average $E_A$ value for blood γ in the range of 30-40 °C has a maximum at Ht40% and is twice as large as $E_A$ for γ plasma and electrolyte solutions (Fig. 6A, Table 2). Due to the formed elements, blood is not a Newtonian fluid and therefore diffusion in it, for example, $Na^+$, does not obey the Stokes-Einstein equation. In addition, the charges on the formed elements, as well as the stratification of the plasma in an electric field, can slow down the diffusion of charges.

A significant difference in the exothermic energy of the solubility of gases $O_2$ ($α_{og}$) and $CO_2$ ($α_{cd}$) in water and plasma is evidenced by the breaks in $F_A$ in the vicinity of $T_h$ and $T_S$ for $O_2$ and their absence for $CO_2$ (Fig. 7). The $E_A$ values for $α_{og}$ in water and plasma in the ranges of 25-36 and 37-40 °C are close to each other (Fig. 7) and practically coincide with the $E_A$ for $γ_H$ in the ranges of 5-35 and 40-60 °C (Table 2). Hence it follows that the dissolution of $O_2$ in IW is associated with HBs rearrangement and the release of $H^+$ from the cell, and the decrease in $E_A$ $α_{og}$ in plasma indicates the effect of proteins on the structure of IhW at $T<T_h$. However, this effect is leveled at $T>T_h$ and the $E_A$ values for $α_{og}$ in plasma and water in the intervals of 0-25 and 25-36 °C practically coincide within the accuracy of measurements in plasma (Fig. 7A).

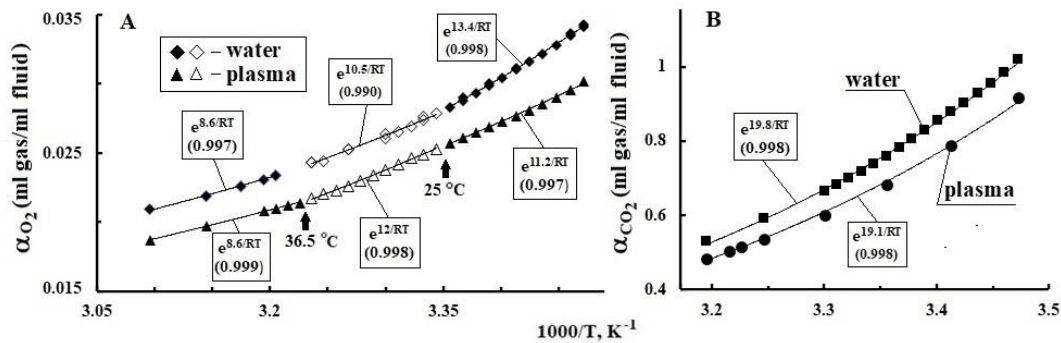

**Fig. 7.** Points – dependences on 1/T solubility in water and plasma of oxygen $α_{og}$ in the range of 15-50 °C (**A**) and carbon dioxide $α_{cd}$ in the range of 15-40 °C (**B**). The lines are $F_A$ approximations. The dependences of the solubilities on T °C are taken from [124].

In the case of carbon dioxide ($α_{cd}$), $E_A$ for water and plasma in the range of 15-40 °C differ slightly and are close to $E_A$ for $T^1$ of water in the range of 0-25 °C, as well as for plasma pH in the range of 20-50 °C (Fig. 7, Table 2). The mechanism of $CO_2$ dissolution is based on the reaction of its fixation by chemical bonds in the $H_2CO_3$ molecule. Since $H^+$ and $OH^-$ are involved in this reaction, the $E_A$ of $CO_2$ dissolution sums up the $E_A$ for pH, as well as the energies of the recombination reactions of $OH^-$ with the electrophilic center of $CO_2$ and $H^+$ with a negative charge on oxygen. The reaction of $CO_2$ fixation proceeds mainly inside the cell; therefore, its energy was not affected by the deformation of HBs by plasma proteins and $E_A$ in water and plasma are almost equal within the accuracy of $α_{cd}$ measurement.

### 4.1.3. Electric factor of water exchange in the parenchyma

Activation of neurons and some brain functions by transcranial stimulation with direct or alternating current or field [125–131] suggests the involvement of $P_w$ and $φ_w$ in brain water homeostasis [132]. $P_w$ and $φ_w$ can polarize diffuse ion fluxes in the ISF and modulate the throughput of water channels of the blood-brain barrier and KVR by the frequencies of the cardiocycle [133, 134]. This modulation may be due to the influence of $φ_w$ on the throughput of transmembrane voltage-gated ion channels and, above all, the $Ca^{2+}$ channel [135-138]. $Ca^{2+}$ is controlled, in particular, by the striction function of pericytes [139-140], which regulate capillary lumens and, hence, the width of KVR gaps [141].

The outer shell of KVR capillaries is mainly formed by the terminal peduncles of astrocytes, which are rich in aquaporin-4 (AQP4) water channels [133, 142, 143]. AQP4, by the principle of electrostatic pumps under the influence of $P_w$ and osmotic pressure, pump water from the capillary segments of the KVR to the ISF and neuropil [144]. In parallel, water can be pumped from LiA into the parenchyma by AQP4 astrocytes of the glial membrane adjacent to the pia mater of the cerebral cortex [145-147]. The parenchyma neuropil also has its own AQP4 network, which is responsible for cellular water exchange and trophism in the mechanism of neurovascular communication [142, 143, 148].

$P_w$ and $V_{ex}$ sensitive elements of cell membranes are ionic and AQP4 channels containing alpha helices and polar molecules [83, 149–151]. The AQP channel consists of six alpha helices and two short helical segments surrounding a water pore in the middle of the channel, which has a d~0.28 nm and a positive charge [152-154]. In AQP alpha helices, the dipole moments of their segments are summed in linear fragments of the helices [83] and, in this case, an electric field with a configuration similar to a multiple thread can be formed in the channel. The lines of force of this field at the entrance to the channel can orientate and impart a rotational-translational motion to the water dipoles, facilitating and accelerating their passage through the pore. The mouths of ion channels of cell membranes and thermoreceptors are covered by a gatekeeper domain, whose structure selectively reacts to neurotransmitter dipoles or ion charges, as well as to local changes in $V_{ex}$ and T [57, 155, 156]. Thus, $P_w$ and $\varphi_w$, through the polarization of water molecules and electromechanical effects in alpha helices, can affect the speed and direction of movement of water molecules along the ion and AQP4 channels in the end legs of astrocytes [144, 147]. A dense network of border and internal channels of AQP4 forms a system of power elements of the hydraulics of the entire parenchyma [144-147]. The electrostatic principle of operation of AQP4 at high distribution densities of capillaries and synapses in the neuropil makes it possible to modulate the water supply system of the entire parenchyma with $P_w$ and $V_{ex}$ oscillations with circadian and heart rhythm frequencies.

### 4.1.4. Effects of chirality and dynamics of protein hydration shells

Blood and CSF can be considered chiral aqueous solutions, since they contain optically active substances that can rotate the plane of polarization of light to the right (D) or left (L). Blood plasma contains D-glucose and proteins albumin, fibrinogen, etc. The dependence of dichroism of proteins and Hb on the configurations of right-handed alpha-helices and T is manifested by changes in the spectra of the ellipticity parameter (θ). An imbalance between the synthesis of beta-amyloid protein (Aβ) and the glymphatic function of the brain leads to the accumulation of amyloid fibrils in the intercellular space of the parenchyma and the development of Alzheimer's disease in elderly people [157-160]. Normal cerebrospinal fluid, along with D-glucose, contains chiral L-lactate, L-glutamate, L-aspartic acid, etc. [161]. Saccharides and proteins have heterogeneous hydration shells (HS) with a thickness depending on the density and strength of their hydrophilic or hydrophobic centers [108]. The hydrophilic bands of molecules are mainly based on OH groups capable of binding up to three water molecules in HS [162]. HS have a strong effect on the structure and functions of biomolecules, but their effect on the dynamics of water around biomolecules is leveled in layers no thicker than 2–3 HB lengths [108, 163–165].

On the other hand, the features of IW thermodynamics in the range of 25-42 °C are manifested in TDs of aqueous solutions of saccharides and proteins simulating FFs. In this range, $F_A$-approximations of the known TDs φ and $[α]_D$ solutions of glucose and other saccharides (Fig. 8, Fig. 9A, Fig. 10A) have $E_A$ values equal to or close to the $E_A$ parameters characterizing the thermodynamics of water at the level of continual rearrangements of the bulk network HBs (n, $C_P$, ε). The $E_A$ values for pure water are determined by van der Waals (up to ~1 kJ/mol) and dipole–dipole interactions (up to ~5 kJ/mol) [100].

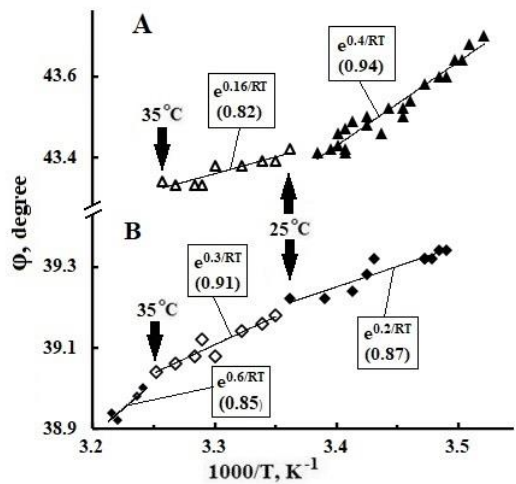

**Fig. 8.** Dependences on 1/T rotation (φ) of 589 nm light of saline solutions (points) and their $F_A$ approximation (lines). **A** - glucose (40%), **B** - dextran (10%). Cuvettes 20 cm. Initial data from [166].

When glucose is dissolved in water, the intensity of the band at 170 cm$^{-1}$ decreases, which corresponds to the HBs extension wave in cooperative tetrahedral domains. In the HS layers closest to glucose, the tetrahedral structure of HBs is distorted, the density of water increases, and the dynamics of HBs rupture slows down by a factor of three [108, 165]. The characteristic time of rearrangement of HBs into HS of biomolecules inside cells is ~27 ps, and the time of exchange of HS with cytoplasmic water is ~4–7 ps, while the lifetime of HB in free water is ~0.5–2 ps [165, 170].

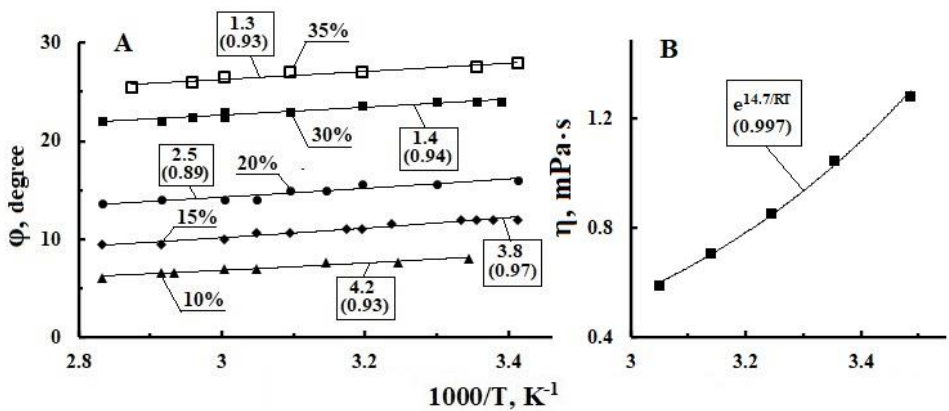

**Fig. 9. A**, dependences on 1/T of the angle of rotation (φ) of 632.8 nm light of aqueous sugar solutions of various concentrations (points) and their $F_A$-approximation (lines). Initial data from [167]. **B**, dependence on 1/T of the viscosity (η) of an aqueous solution of levoglucosan (100 g/L). Initial data [168].

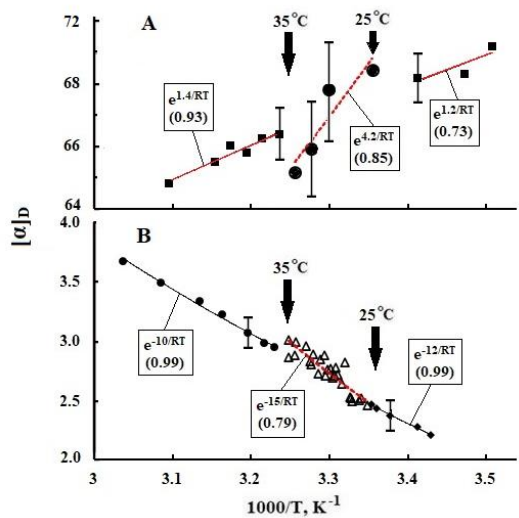

**Fig. 10.** Dependences on 1/T of specific optical rotation $[\alpha]_D$ (deg dm$^{-1}$ cm$^3$ g$^{-1}$) (points) aqueous solutions. **A**, of allyl lactoside (0.2 mol/l), **B**, levoglucosan (0.1 mol/l). The lines are $F_A$ approximations. Initial data from [169]

Unlike flat glucose with 5 OH groups, levoglucosan has a three-dimensional structure and only 3 OH groups. With this structure, the binding of levoglucosan to three water molecules in solution results in a decrease its $[\alpha]_D$ by more than an order of magnitude relative to the $[\alpha]_D$ of glucose solution [169] (Fig. 8 and Fig. 10B). The rotational dynamics of the chiral centers of levoglucosan and the dipoles of free water molecules are compared, as evidenced by the closeness of ~15 kJ/mol in the range of 25-37 °C of the $E_A$ values for $[\alpha]_D$ and η levoglucosan (Fig. 10B, [168]) to the $E_A$ values for η and $\tau_D$ of bulk water (Table 1). Disaccharides have 7-8 OH groups and their HS, like glucose in the range of 25-37 °C, ensures the integration of the rotational dynamics of the chiral centers of molecules into the cooperative dynamics of HBs. This is confirmed by the correlation of the $E_A$ values for φ and $[\alpha]_D$ saccharides (Fig. 8, Fig. 9 and 10A) with the $E_A$ of the characteristic parameters of dipole domain rearrangements and the HBs structure of water (n, ε, $C_P$).

In addition to monosaccharides and polysaccharides, the chiral and dynamic properties of ISF and blood plasma can be affected by the chirality and mobility of alpha helices that make up blood and CSF proteins. The effect of anomalies in water thermodynamics on the dynamics and chirality of proteins is manifested by breaks in $F_A$ and drops in $E_A$ at points $T_h$ and $T_b$ on TDs θ and $R_H$ of model solutions of albumin and Hb (Fig. 11, Fig. 12A) and lysozyme [107]. The dependence of $E_A$ rearrangement of alpha helix conformations on water dynamics in the hydration shell of the protein is confirmed by kinks in $F_A$ close to $T_b$ on TDs content (%) and elasticity ($u^2$) of alpha helices in Hb (Fig. 11C and Fig. 13B). The U-shape of the TD θ-parameter of albumin is due to variations in the contributions to the light absorption of its two alpha-helices (Fig. 11A) [171]. $E_A$ values for θ and $u^2$ fragments of mammalian Hb alpha helices and Aβ amyloid proteins vary in different T-ranges from ~6 to 14 kJ/mol (Fig. 11, Fig. 12D and Fig. 13B) and correlate with the $E_A$ range for $T^1$ in HS proteins (~8-13 kJ/mol) [109, 166]. It follows that the local transformation of protein alpha-helices is mainly limited by the dynamics of HBs within HS, which screens the effects of HBs in bulk water [108, 172], including an increase in $H^+$ concentration by 4 orders of magnitude (Fig. 12D). On the other hand, estimates of the $E_A$ of mammalian Hb transformations based on the TDs of the $R_H$ parameter give values of 16÷26 kJ/mol at T>35 °C (Fig. 13A), which correlate with the $E_A$ values characteristic of the dynamic parameters of bulk water (η, D, $\tau_D$), and the $E_A$ values at T<35 °C (3÷7 kJ/mol) correlate with the $E_A$ for ε water (Table 1) and the specific heat of crystallization of water (~6 kJ/mol).

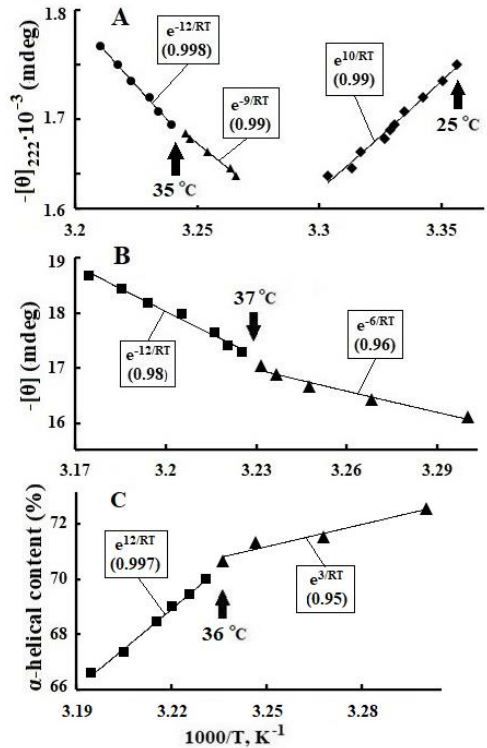

**Fig. 11.** Points - dependences on 1/T of the angle of ellipticity ($[θ]_{222}$) of aqueous solutions: **A**, albumin (0.5 g/l), **B**, human oxyhemoglobin and **C**, the content (%) of alpha helices in it. The lines are $F_A$ approximations. Initial data for **A** from [172] and for **B**, **C** from [173].

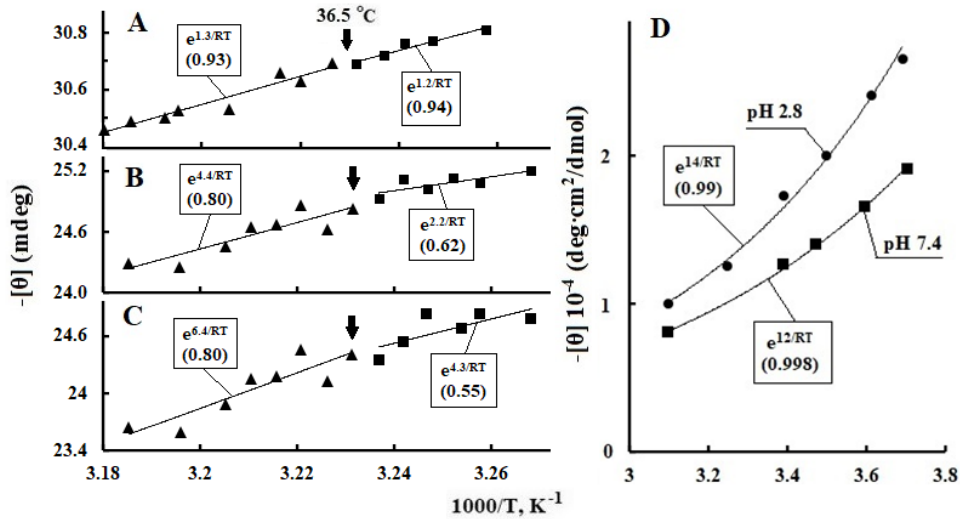

**Fig. 12.** Points – dependences on 1/T of the angle of ellipticity ([θ]) of aqueous solutions of hemoglobins (0.1 g/l): **A**, heron; **B**, pig; **C**, camel. Lines are $F_A$-approximations. Initial data from [19]. **D**, dependence of helix formation for amyloid P-peptides on 1/T, at pH 2.8 and pH 7.4 (points). Lines – $F_A$ approximations. Initial data from [157].

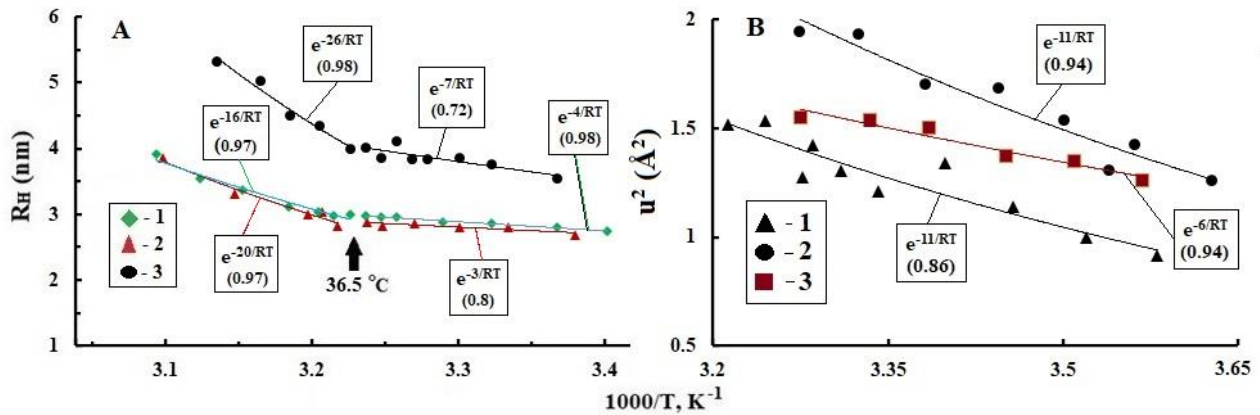

**Fig. 13. A**, dependences on 1/T $R_H$ of aqueous solutions of hemoglobins (0.1 g/l) (points): **1** - human, **2** - bull and **3** – platypus; **B**, mean displacement squares ($u^2$) of hemoglobin solutions (1 g/ml): **1** – human, **2** – bull, **3** – chicken. The lines are $F_A$ approximations. Initial data from [19, 174].

In birds, the average values of $E_A$ for $u^2$ and θ of Hb solution over T-intervals are 2÷5 times lower than $E_A$ for mammals (Fig. 12A and 13B). Hence, it follows that the alpha-helices of the Hb protein globule of birds have a loose HS and practically do not react to the rearrangement of HBs of bulk water in the vicinity of the $T_S$ (Fig. 12A). This result is consistent with the genetic features of the anatomy and physiology of the circulatory and CSF systems of the brain of birds. Erythrocytes in birds are much larger than in mammals and have nuclei, albumin in their blood is ~1.7 times less, and $T_b$ is ~4 °C higher than in humans and mice. In female birds, the right ovary is atrophied; in males, the erection of the penis is due to the influx of lymph into it [175]. Birds practically there is no neocortex and therefore no need for a glymphatic system.

Mammalian erythrocytes with d~8 µm and a thickness of 1÷2 µm, due to the elasticity of the plasma membrane in vivo, can reversibly change shape and pass through capillaries with d~3 µm. The viscosity of the erythrocyte cytoplasm at $T_S$ is ~3 times higher than the plasma viscosity and is a Newtonian fluid [178]. Up to ~70% of the erythrocyte volume is occupied by water and ~85-90% of it is in the free state, and ~10-15% in HS [170, 179]. In experiments *in vitro*, erythrocytes under pressure, acquiring the shape of a spindle, pass through holes d~1.3–2 µm [58, 176, 180]. Transformation of an erythrocyte in the range of 35-38°C is accompanied by Hb aggregation [58, 181] with a 1.5-fold decrease in the erythrocyte volume due to the release of up to 50% of cytosolic water through AQP1 in the plasmalemma [176].

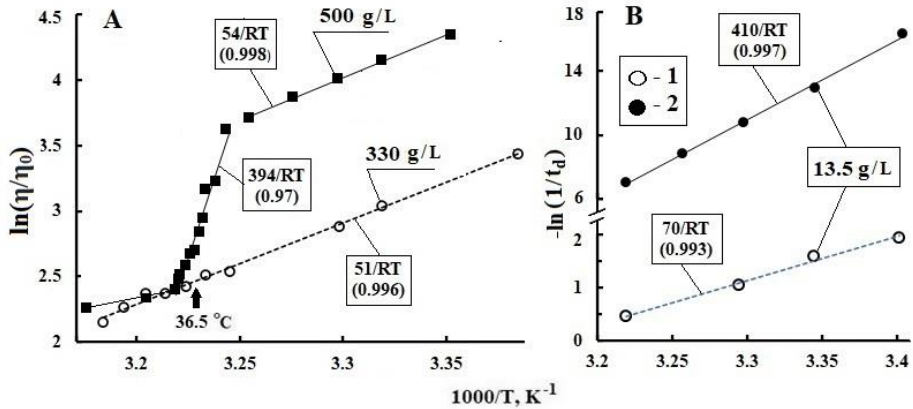

**Fig. 14. A**, dependence on 1/T relative shear viscosity η/η₀ of the hemoglobin solutions at various concentrations (η₀ – viscosity of the solvent). The lines are $F_A$ approximations. Modified from [176]. **B**, dependence on 1/T of the rate of aggregation (**1**) and crystallization (**2**) of oxyhemoglobin C in an aqueous solution at pH 7.4. Modified from [177].

In the vicinity of $T_S$, weak breaks in $F_A$ on TDs θ and $R_H$ of model Hb solutions (0.1 g/L) of humans and mammals are revealed (Fig. 12A, Fig. 13A). On TDs η of human Hb model solutions with a concentration of 450 and 510 g/L, when heated from 35 to ~38 °C, a sharp drop in shear η is observed from $E_A$ ~200 and 394 kJ/mol against the background of a gradual decrease in η from $E_A$~54 kJ/mol at T<35 °C. There is no such decrease in the value of η when heating a saline solution with a normal concentration of Hb (330 g/L) and η gradually decreases in the range of 22-41 °C from $E_A$~51 kJ/mol (Fig. 14A).

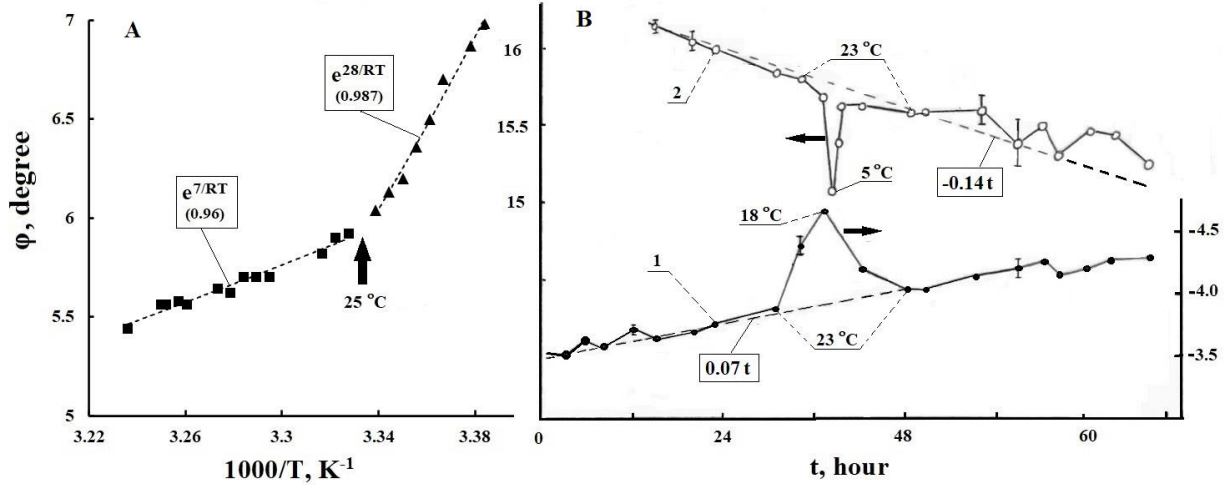

**Fig. 15.** Points - dependence on 1/T of the angle of rotation ($\varphi$) of light 589 nm. **A**, saline gelatin (4%), cuvette 200 mm, dotted line $F_A$ approximations; **B**, saline gelatin (4%) (**1**), saline gelatin (2%) + sugar (10%) (**2**), cuvettes 100 mm. Dashed lines are curve slopes. At ~36 hours, both cooled reversibly from 23 to 18°C and 5°C. Figures adapted from [166, 184].

It is known [177, 182] that the $E_A$ values of the process of aggregation and nucleatization of Hb and proteins in solutions are of the order of ~70 kJ/mol, while the $E_A$ of crystallization, depending on the concentration and T, varies in the range of 150–410 kJ/mol [182, 183]. The closeness of the $E_A$ TDs η values to these values in the range of 35–38 °C for model solutions with Hb concentrations >330 g/L indicates the effect of HBs rearrangement in the vicinity of $T_S$ on the dynamics of protein aggregation, crystallization, and fibrillation.

The intercellular space of the parenchyma is a highly hydrated mesh medium, the elastic skeleton of which consists of hyaluronic acid, collagens and elastin. ISF moves in channel-connected microspaces of the gel structure formed from collagen (90%) and other (10%) proteins [72]. A gel-like medium is, in principle, simulated by saline solutions based on gelatin and sugar (Fig. 15). The value of $E_A$=28 kJ/mol for $\varphi$ gelatin (4%) at T<$T_h$ (Fig. 15A) is close to $E_A$ for $D_w$ and $\tau_D$ of supercooled water in the range from 0 to -20 °C [86]. At T>$T_h$, the value of $E_A$ for $\varphi$ gelatin solution corresponds to $E_A$ for θ alpha-helices in aqueous solutions of Hb (Fig. 11). These correlations illustrate the differences in water dynamics in the ISF in the IhW and IW states. On Fig. 15B shows that at T=23°C, the chirality (-$\varphi$) of jelly-like gelatin (4%) grows with time twice as slowly as $\varphi$ of more mobile gelatin (2%), in which the process of jelly formation is not completed. With a decrease in T, -$\varphi$ gelatin reversibly increases for a short time and $\varphi$ of the sugar solution decreases by this amount. Hence it follows that the chirality of ISF can be represented by the algebraic sum of the chirality D and L of the metabolites.

### 4.2. CSF circulation in the parenchyma during wakefulness

Due to the anatomy of the skull and the incompressibility of the aquatic environment of the brain, intracranial pressure fluctuations caused by pulse waves in the arteries are focused on the CSF of the lateral and third ventricles. Response pressure waves CSF in the ventricles with a frequency of ~1 Hz diverge centrifugally through the brain tissues adjacent to the ventricles and, having passed through the system of connecting channels and the fourth ventricle, diverge through the cisterns of the arachnoid spaces of the spinal cord and brain, fading in LiA [148, 185-186]. Together with these fluctuations, centripetal respiratory oscillations CSF with a frequency of 0.3 Hz are observed in the near-venous areas of the brain. In [187], similarly to [67], using computer programs for signal filtering and a $V_R$-wave as a reference, a third type of CSF dynamics pulsation with frequencies of 0.001–0.73 Hz and a sporadic spatiotemporal distribution in brain [17]. The frequencies of these oscillations are related to the noise activity of the brain [188, 189], its EEG frequency spectrum may include EEG potentials related to the electrophysics of neurovascular communication [190]. The frequencies of these potentials will correspond to changes in the dynamics of ISF microcirculation and hyperemia of local areas of the parenchyma, synchronous with the modulation of pulse oscillations $\varphi_w$ by B-wave frequencies and the rhythm of precapillary sphincters [191-193].

In the waking state, the continuous water supply of the parenchyma metabolism can go through the following channels. Water from arterial blood enters the capillary segment of the KVR system through blood-brain barrier channels with the participation of AQP4, where it mixes with water from LiA, which enters through arterial KVR (perivascular pumping) [194-197]. Note that the diffusion of solutes in gel-like ISF at $T_b$ decreases by ~30–80% compared to diffusion in pure water [27, 73, 74, 198, 199]. At the same time, the inflow of water into the parenchyma from arterial blood capillaries and a weak inflow from LiA along the KVR and AQP4 channels of glial membrane astrocytes are normally equal to the outflow of water through the paravascular spaces of the draining veins [199, 200]. The hydraulics of drainage of water and waste into venules, similar to the mechanism of $CO_2$ drainage from ISF, seems to be provided by the difference in hydrostatic and osmotic pressures between the ISF and the blood of postcapillary venules [193].

Thus, in the waking state, the total hydraulics of the water channels of the capillaries and the system of border and internal channels AQP4 of the parenchyma can ensure the functional autonomy of the ISF microcirculation against the background of pulsed fluctuations in the hydrostatic pressure CSF [146]. The relationship of blood dynamics (Bld), CSF and LiA with the circulation of ISF in the parenchyma was represented by a diagram of communicating channels (4.1). In (4.1), the molecular mechanisms of water exchange between ISF and LiA (2-channel and -2-channel reverse), as well as between ISF and Bld (3-channel arterial, -3-channel venous), have not been fully elucidated [27, 195, 197]. 4-channel – secretion of CSF from arterial blood in the choroid plexuses of the lateral ventricles, -4-channel – reabsorption of LiA by venous blood through the granulation of the arachnoid of the superior sagittal sinus.

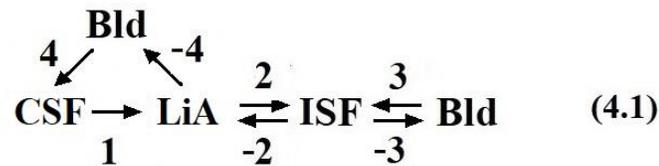

$$\begin{array}{c} \text{Bld} \\ {}^{4}\swarrow \quad \nwarrow {}^{-4} \\ \text{CSF} \xrightarrow{} \text{LiA} \underset{-2}{\overset{2}{\rightleftarrows}} \text{ISF} \underset{-3}{\overset{3}{\rightleftarrows}} \text{Bld} \\ {}_{1} \end{array} \qquad (4.1)$$

### 4.3. Circadian rhythm and the glymphatic system of the brain
### 4.3.1. Circadian factors of day and night

The global biogenic factors of the circadian rhythm are sunlight during the day and cold at night. Their influence caused the emergence and development of receptors in animals that react to visible light and cold. In mammals, these receptors are localized in the eyeball in combination with the vitreous body (VB), which consists of ~99% water. The ganglionic layer of the photosensitive retina is adjacent to the dorsal surface of the VB, and the thermoreceptors are concentrated in the cornea, which has close thermal contact with the VB through the lens and aqueous humor of the eye chambers, which is similar in composition and functions to blood plasma. The eyeball is thermally isolated from the bones of the orbit by a layer of adipose tissue and from the external environment by eyelids, the fiber of which is devoid of a fatty layer. Therefore, the cornea and VB in sleep with the eyes closed for eyelids retain adequate thermal contact with the external environment, and the average temperature of the cornea and VB is $T_w \pm 0.5$ °C [201]. Thermal insulation of brain tissues by the cranial bone and skin ensures their T stabilization during sleep at 36.5 °C. Apparently, such temperature deviations of the eyeball and brain from $T_w$ provide high sensitivity of corneal thermoreceptors to changes in external T and activation of the glymphatic system of the brain during falling asleep (Section 4.1.1.).

The gel-like substance VB within its own fibrillar shell is "reinforced" with threads of collagen and hyaluronic acid. During the transition from wakefulness to sleep, the rate of aqueous humor flow in VB decreases by almost half [201]. At a qualitative level, the circadian rhythm manifests itself in intraocular pressure [202], and in the phase of sleep with rapid eye movement (REM phase) intraocular pressure fluctuations were minimal, while in the NREM phase they reached a maximum in spindle oscillations [55, 203]. These oscillations will modulate the movement of the ocular fluid along the VB into the perivascular space of the optic nerve [204], simultaneously participating, together with capillary water, in retinal clearance in the state of sleep or drowsiness.

Cells of the ganglion layer convert light information into optic nerve impulses, which, after primary processing in the thalamus, are fanned out to the visual area of the cerebral cortex [57]. At the same time, ~1-2% of cells in the ganglionic layer are considered special (ipRGC), they synchronize their physiology and activity with the light-dark circadian cycle [205]. The photopigment of ipRGC is melanopsin and they are directly associated with the suprachiasmatic nucleus (SCN) of the hypothalamus [206-209]. Apparently, the signaling relationship between ipRGC and SCN plays a key role in triggering the mechanism of switching brain homeostasis from daytime to nighttime mode [25, 210].

Taking into account the phylogenetic synergy of dark and cold factors, it can be assumed that thermoreceptors in the cornea of the eye have retained their contribution to the control of nocturnal brain metabolism. The density of trigeminal nerve endings that respond to heat (pain) and cold in the cornea of the human eye is two orders of magnitude higher than in the skin of the fingers [211-216]. The high sensitivity of cold receptors in the cornea is due to membrane voltage-dependent cation channels TRPM8 [212–216], which have a protein structure similar to AQP4 (see Section 4.3.1). In humans and terrestrial mammals, the signaling systems of light and cold receptors can, mediated by the functions of the thalamus, hypothalamus, pineal gland, and brainstem structures, provide a harmonious combination of two brain metabolic regimes corresponding to wakefulness and sleep [28, 54, 217-224].

In mammals, upon falling asleep, the output neurohumoral signals of the SCN activate the metabolism of the pineal gland [25, 28, 209, 210, 222, 223]. The content of melatonin, serotonin, norepinephrine and other neurotransmitters responsible for switching homeostasis and hydrodynamics of brain FFs to the mode of glymphatic function increases in the blood and cerebrospinal fluid. Apparently, the duration of nocturnal sleep in terrestrial mammals correlates with the level of need for cleaning the parenchyma of the cortex and replenishing the trophic-communicative resources of the brain. Genetically, this need reflects the specifics of the lifestyle and physical characteristics of the animal's habitat. Accordingly, in mammals, variations in the duration of nocturnal sleep correlate with the physiological parameters of the key elements of the circadian rhythm control systems, which include the photo- and thermoreceptors of the eye, the SCN, and the pineal gland [205, 209]. The pineal gland is absent in the electric skate, crocodile, cetaceans, is not found in the dolphin, and is very small in the elephant [225]. In mammals of the northern latitudes, the epiphysis is larger than in the inhabitants of the southern latitudes. The average volume of the epiphysis (in $mm^3$): 6÷12 (predators); 60÷300 (ungulates); ~180 (monkey); ~200 (human) [225]. Similarly, most terrestrial mammals show a direct correlation between sleep duration and their eyeball diameter ($d_e$) (Fig. 16). The value of $d_e$ determines the volume of VB, the area of the cornea, and the number of ipRGC cells in the corneal fiber layer. The absence of the epiphysis and the REM phase in dolphins and cetaceans [226] may be due to the lack of expression of diurnal fluctuations in world ocean T or the leveling of the difference in the effect of the biogenic solar factor on the biosphere day and night by the aquatic environment [28, 29].

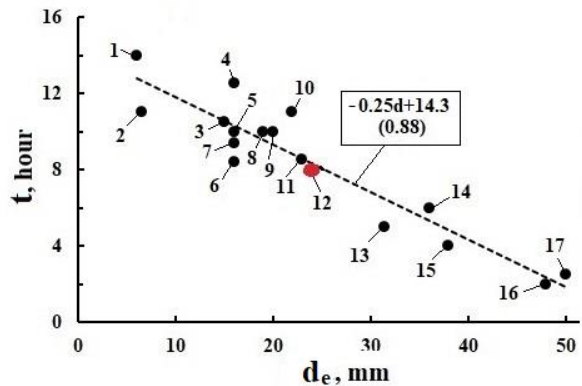

**Fig. 16.** Dependence of the duration of sleep per day (t) of mammals on the diameter of their eyeball ($d_e$): 1 - platypus, 2 - rat, 3 - adult squirrel, 4 - cat, 5 - fox, 6 - rabbit, 7 - lemur mouse, 8 - rhesus monkey, 9 - dog, 10 - orangutan, 11 - pig, 12 - Human, 13 - goat, 14 - sea lion, 15 - elephant, 16 - giraffe, 17 - horse. Open data on GOOGLE.

**4.3.2. Two regimes of the glymphatic system**

Normal EEG during sleep is divided into NREM and REM phases [54, 190, 227, 228], the physiology of which corresponds to two conditional modes of operation of the glymphatic system (GS), electrochemical (GS1) and dynamic (GS2) [25-27]. When falling asleep in the GS1 regime, different stages of NREM sleep dominate, and the eye muscles and neuropil are periodically replenished with glucose for 15–20 min [161, 218, 229]. At the same time, the level of consumption of glucose and oxygen in the parenchyma does not change significantly [120, 230,

231]. In the GS1 regime, relaxation processes of synaptic plasticity and chemical neutralization of toxins occur with the participation of water and melatonin [108, 163, 220, 232, 233]. At the same time, a decrease in blood flow in the brain by ~25% and blood volume by ~10% is accompanied by an influx of CSF to the third and fourth ventricles [221, 234]. It can be assumed that a decrease in the temperature of the brain and VB in the NREM phase [217, 228], as well as an increase in $CO_2$ and acidity in the ISF [25, 144], initiate switching from the GS1 regime to the GS2 regime (REM sleep).

GS2 is characterized by rapid eye movement, dreams, and, unlike GS1, a sharp increase in cerebral blood flow with dilation of arterioles and capillaries [190, 227]. Synesthesia of vision with almost all somatosensory [57, 219] suggests convergence of the nervous systems of thermoreceptors and oculomotor muscles [54, 235, 236]. Rapid eye movement enhances the signaling of thermoreceptors in the cornea and cells of the corneal layer of the retina along nerve connections with the thalamus, hypothalamus, and midbrain nuclei [224]. An increase in blood flow intensifies water exchange, which is necessary for flushing out toxins from the parenchyma into venules through the -3 channel of scheme (4.1) [120, 161, 190, 227]. The expansion of the lumens of arterioles and capillaries enhances the polarization effects $P_w$ and $\varphi_w$ in the electrophysics of synaptic plasticity (see Section 1.3). The duration of GS2 seems to be determined by the time when glucose is depleted in the oculomotor muscles and the threshold value of lactate is reached in them [161].

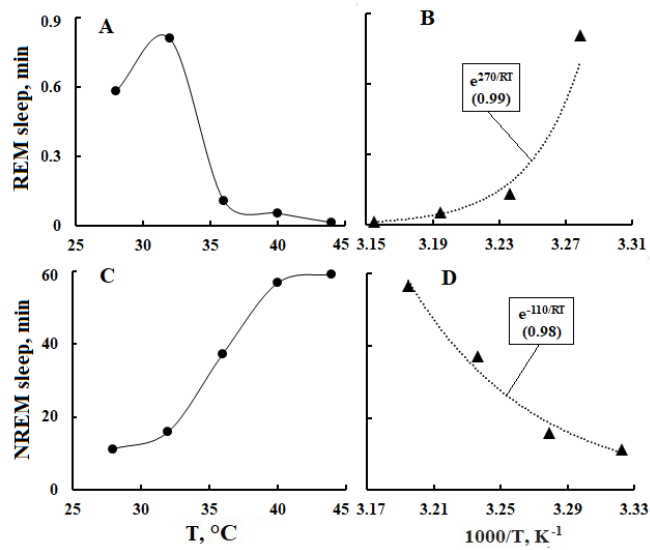

**Fig. 17.** Points - dependence of the duration of sleep phases REM (**A**, **B**) and NREM (**C**, **D**) of mice on T (**A**, **C**) and 1/T (**B**, **D**) in the period of 23.00-01.00 hours; lines are envelopes (**A**, **B**) and $F_A$ approximations (**B**, **D**). Initial data from [237].

In the signaling systems of the GS1 and GS2 regimes, the key role in the range of 32–40°C is played by the thermodynamics of FFs and hydration shells of protein channels AQP4, TRPM8, and other voltage-gated ion channels. Their physiological specialization is manifested in a significant difference in the duration and $E_A$ of NREM and REM sleep phases in mice (Fig. 17). The effective $E_A$=110 kJ/mol response that determines the duration of the NREM sleep phase (Fig. 17D) almost coincides with the $E_A$=112 kJ/mol electrical stimulation of rat retinal ganglion cells, which follows from the $F_A$ approximation of TD of their threshold current obtained in vitro (24.4 °C - 239 µA, 29.8 °C - 101 µA, 33.8 °C - 60 µA, $R^2$=0.999) [238]. On the other hand, the effective $E_A$=270 kJ/mol of a similar REM sleep phase response (Fig. 17B) appears to be of the same order of magnitude as the $E_A$ of TRPM8 threshold currents (Fig. 18).

### 4.3.3. Thermodynamics of the glymphatic system

The $T_S$ value of ~36.5 °C during sleep is characteristic of fluids – ISF, LiA, and also CSF of the third and lateral ventricles of the brain [24-27, 239]. The SCN and pineal gland will have a temperature close to $T_S$, since they are adjacent to the third ventricle of the brain [206, 240]. At external T=24±1.1 °C, the temperature under the tongue is 36.6±0.5 °C [201]. On the surface of the cornea, due to heat exchange with the external environment, T=34.7±1.1 °C is established, which coincides with $T_w$ and with the threshold T of the operation of the TRPM8 cold channel [212]. In the VB volume T=33.9±0.4 °C [201, 241], and in the ganglionic layer of the retina it is 34.8÷35.2 °C [241]. Thus, thanks to VB, the eyeball can serve as a temperature sensor that regulates the operation mode of TRPM8 channels of corneal nerves and voltage-dependent $K^+$ channels of ganglion cell axons, including ipRGC, in accordance with external T [207, 208].

In the sleep state T of the cornea and VB are below $T_w$ and the thermodynamics of their fluids should be dominated by clustering processes of HBs water (Section 4.1.1). These processes in FFs and in the structures of hydration shells of TRPM8 proteins and VB fibrils will initiate the processes of aggregation and crystallization [108, 182]. Indeed, the values of the effective energies of cold activation of TRPM8 (Fig. 17B, Fig. 18A) and generation of ionic currents (Fig. 18B, 18C and Fig. 19A) are comparable to the $E_A$ of the processes of aggregation and crystallization of alpha helices in high concentration Hb solutions (Fig. 14). The influence of HS dynamics on TRPM8 currents $E_A$ is evidenced by the doubling of $E_A$ with the addition of menthol, a specific TRPM8 activator, to the bath solution (Fig. 18C, [108, 156, 213]) and a 1.5-fold decrease in $E_A$ current with the addition of a cold receptor blocker, BCTC (Fig. 18B, [212]). The OH group of menthol and the hydrophilic centers of BCTC, when interacting with the protein domains of the TRPM8 and TRPV1 channels, initiate the redistribution of charges between amino acids and cations in them [242-244]. In this case, local and transmembrane potentials arise [150, 151, 213], which control the opening-closing mechanisms of the channels and ensure the passage of cations through them into or out of the neuron [212, 213]. The linear dependence of the TRPM8 EA currents on the potential difference follows from the comparison of the TDs currents in Fig. 18C and Fig. 19A.

Corneal thermoreceptors with TRPV1 channels have a maximum sensitivity at T~40 °C [245] close to the pain threshold of 42 °C [32, 57, 219]. In the range from $T_h$ to ~40 °C, the $E_A$ values of the ion current in the TRPV1 channel and in the bath buffer solution is 21 kJ/mol (Fig. 19B) and 17 kJ/mol [246], respectively. These $E_A$ values correlate with the $E_A$ of the rotational-translational diffusion of water ($\tau_D$, $D_w$, Table 1) and electrolytes ($T^1$, Table 2) in the respective T ranges. The increase in the $E_A$ ion current in TRPV1 at T>40 °C to 48 kJ/mol is apparently due to the specifics of the molecular mechanism of the TRPV1 response to T above the pain threshold [57, 245]. The average current $E_A$ value in TRPV1 in the range of 33-42 °C is ~34 kJ/mol, which coincides with the EA (33 kJ/mol) TD of the saltatoric impulse transmission rate in the myelinated fiber (Fig. 19C).

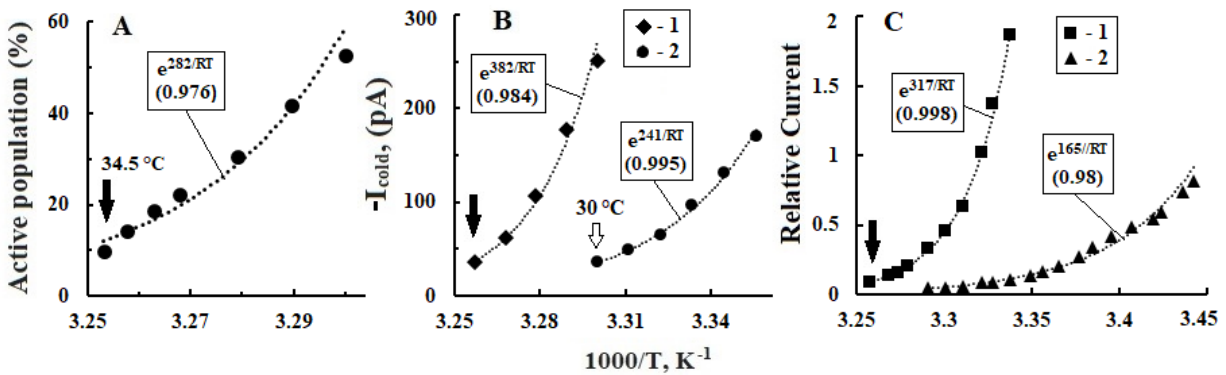

**Fig. 18.** Points – dependences on 1/T of the bath. **A**, Cumulative distribution of active trigeminal neurons with TRPM8 channels. **B**, current in the TRPM8 channels of the trigeminal nerve (**1**) and the effect of on it of the addition of 1 µM BCTC (**2**). **C**, relative current in TRPM8 channels (**2**) at -60 mV and influence on it of 30 µM menthol (**1**). The arrows mark the threshold cold T. The lines are $F_A$ approximations. Initial data **A** from [243], **B** from [212], **C** from [244].

The consistency of $E_A$ values confirms the dependence of the kinetics of the saltatory mechanism on the efficiency of the operation of voltage-dependent $K^+$ channels in the nodes of Ranvier [28]. The saltatory mechanism still acts on fibers with a removed myelin sheath at T<36.5

°C, but is blocked at T>36.5 °C [247]. The blocking can be associated with the destruction at T>36.5 °C of the HBs structure in the axoplasm, which is necessary for the generation and propagation between the nodes of polarization waves that activate $K^+$ channels in them [28]. It also follows from this that during sleep at T ~ 35-36.5 °C, the rate of AP transmission to the SCN along the ipRGC axons, which do not have myelin sheaths within the retina [248], will be about 50 m/s (Fig. 19C). Let us assume that activation of the retina by light upon awakening is associated with an increase in T axons of ganglion cells outside the retina to $T_b$, and in the case of ipRGC this leads to blocking of their communication with the SCN [247]. Thus, temperature regulation of the circadian rhythm can occur.

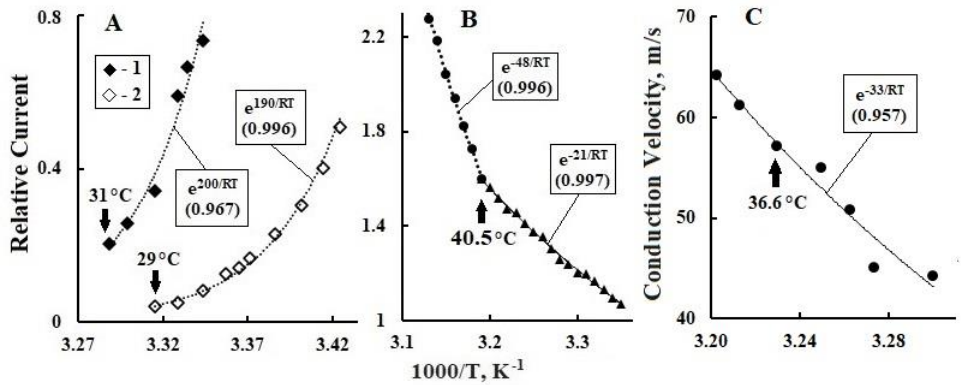

**Fig. 19.** Dependences on 1/T of relative currents in various ion channels (points). The lines are $F_A$ approximations. **A**, the TRPM8 channel of whole cells at a transmembrane potential of +100 (**1**) and -80 (**2**) mV, the arrows indicate the threshold cold T. **B**, the TRPV1 channel, the arrow marks the temperature of maximum sensitivity to heat. **C**, speed of action potential propagation between nodes of Ranvier of the myelinated fiber. Initial dependences on T: **A** from [244], **B** from [245], **C** from [247].

In mice, it was established by iontophoresis that during falling asleep and during anesthesia, the rate of diffusion of the tetramethylammonium cation in ISF increases by 60% [25, 26, 70, 73-249]. It is known that 30 min after anesthesia induction, T in mice decreases from ~37°C to ~24°C [219]. With this in mind, it can be assumed that in the $T_S$-$T_w$ range, the gel-sol transition in ISF initiates the decomposition of HBs clusters (Section 4.1.1), which leads to an increase in $D_w$ and ion mobility by 60%. At the same time, narrowing of the lumen of blood vessels under the influence of serotonin and norepinephrine produced by the pineal gland in the GS1 regime [25, 211] will lead to a widening of the KVR channels and an increase in the influx of LiA into the parenchyma. Similar processes are manifested by kinks in $F_A$ approximations of TDs $\gamma_w$, $\gamma_H$ and $\gamma$, $\alpha_{og}$ of water and plasma (Fig. 6 and Fig. 7A, Table 2), as well as $\alpha_D$, $\theta$ and $R_H$ of model solutions of human and mammalian plasma and cerebrospinal fluid (Fig. 11, Fig. 12, Fig. 13).

## 5. Conclusion

A systematic analysis of the known temperature dependences of the kinetic characteristics of the signaling and trophic functions of the brain, carried out in this work, showed that the electrical and dynamic properties of water play a key role in their molecular mechanisms. This is confirmed by the correlations between the activation energies of the rearrangements of hydrogen bonds in water and the temperature dependences of the physicochemical parameters of physiological fluids. The cooperation of water dipoles into domains contributes to the polarization of the near-wall plasma layers in the blood vessels. The estimates of ECG signal amplitudes in the EEG spectrum indicate that the potentials of water domains in the plasma of arterioles and capillaries can modulate the conductance of ion and water channels of the blood-brain barrier and astrocyte membranes by the frequencies of the cardiocycle. The synergy of thermal libration fluctuations of the water molecule and exothermic proton hops minimizes the heat capacity of water at ~34.5 °C and stabilizes the thermodynamics of human eyeball and brain fluids in the range of 33–40 °C. In the process of phylogenesis in the physiology of terrestrial mammals, a mechanism of adaptation to the circadian rhythm has developed, taking into account the specifics of the lifestyle and the physical characteristics of the habitat. This mechanism is based on the integration of the visual system with almost all somatosensory systems and the localization of light and cold receptors in the eyeball, which are responsible for switching the functions of the pineal gland and the suprachiasmatic nucleus from daytime to nighttime mode. This is confirmed by correlations between the dependences of sleep duration in most mammals on the diameter of the eyeball and the mass of the pineal gland. In addition, there is a strict agreement between the nighttime temperature of the eyeball and the triggering threshold of the TRPM8 channel of the cold receptor of the cornea, as well as the daytime temperature of the brain with the threshold for blocking the communication channel of retinal ganglion cells ipRGC with the suprachiasmatic nucleus. The neural connections of the eyeball with the brain seem to be involved in the division of the nocturnal brain metabolism into two phases of sleep – NREM and REM and into two regimes of the glymphatic system of the brain – electrochemical and dynamic. The first regime is characterized by relaxation processes of synaptic plasticity and chemical neutralization of toxins with the participation of water and melatonin. Activation of the oculomotor muscles and a sharp increase in cerebral blood flow in the second regime intensify water exchange in the parenchyma and flush out toxins into the venous system of the brain. In both regimes, oscillations of the polarization potentials of arterioles and capillaries of the parenchyma can play a significant role.

From the results of the work, it follows that classical electrophysics and thermodynamics of water underlie the neurophysiology of the basic functions of the brain, which are characteristic of all mammals, including humans. It can be hoped that deepening the knowledge of the properties of water to the level of subquantum physics will make it possible to study the nature of the physical uniqueness of the human mind [28, 29, 250, 251].